%% file: paper.tex
\documentclass[]{spectral_ai}

\usepackage[toc,page,header]{appendix}

\usepackage{wrapfig}
\usepackage{multirow}
\usepackage{makecell}
\usepackage{colortbl}
\usepackage{amsfonts}
\usepackage{amsmath}
\usepackage{float}
\usepackage{seqsplit}
\definecolor{colorfirst}{rgb}{.866,.945, 0.831} 
\definecolor{colorsecond}{rgb}{1, 0.98, 0.83} 
\definecolor{colorthird}{rgb}{0.76, 0.87, 0.92} 
\definecolor{colorcite}{rgb}{0.212, 0.490, 0.741}

\title{A large-scale foundation model enables simulation-to-real adaptation for nuclear magnetic resonance-based molecular structure analysis}

\author[1]{Chen Yang}
\author[1]{Zheng Fang}
\author[2]{Hanyu Sun}
\author[3]{Fanjie Xu}
\author[4]{Hongxin Xiang}
\author[5]{Hanyu Gao}
\author[4]{Xiangxiang Zeng}
\author[6]{Yuqiang Li$^{\dagger}$}
\author[2]{Xiaojian Wang$^{\dagger}$}
\author[1, 5]{Jun Xia$^{\dagger}$}
\affiliation[1]{The Hong Kong University of Science and Technology (Guangzhou)}
\affiliation[2]{Peking Union Medical College and Chinese Academy of Medical Sciences}
\affiliation[3]{Xiamen University}
\affiliation[4]{Hunan University}
\affiliation[5]{The Hong Kong University of Science and Technology}
\affiliation[6]{Shanghai Artificial Intelligence Laboratory}
\abstract{
Nuclear Magnetic Resonance (NMR) spectroscopy is a powerful tool for molecular structure analysis, and spectral artificial intelligence offers great potential for its rapid and automated interpretation. However, the scarcity of experimental NMR datasets has constrained deep learning in this domain to narrow, task-specific applications that lack broad generalization.
Here, we introduce UltraNMR, a large-scale foundation model for NMR that leverages the intrinsic properties of NMR spectra to learn generalizable spectral representations. We collected 158 million paired simulated $^{1}$H and $^{13}$C NMR spectra to train UltraNMR, employing multiple domain-specific pre-training objectives. UltraNMR captures both intra-spectral and inter-spectral dependencies, enabling seamless simulation-to-real adaptation.
We demonstrate that adapting UltraNMR to a range of molecular structure analysis tasks on experimental NMR spectra consistently yields state-of-the-art performance and clearly outperforms UltraNMR variants trained directly on downstream data without simulation pre-training. 
We also construct a large-scale NMR spectral vector library by encoding simulated NMR spectra using UltraNMR, covering 94 million unique molecules and enabling effective structure-aware retrieval. In real-world applications, UltraNMR facilitates the structural elucidation of two previously unknown natural products from Chinese herbal medicines recorded in the Chinese Pharmacopoeia. These results suggest that large-scale simulation pre-training can effectively bridge the simulation-to-real gap, enabling robust and generalizable molecular structure analysis of real-world NMR spectra.
}

\correspondence{Yuqiang Li (\href{mailto:liyuqiang@pjlab.org.cn}{liyuqiang@pjlab.org.cn});
Xiaojian Wang (\href{mailto:wangxiaojian@imm.ac.cn}{wangxiaojian@imm.ac.cn});
Jun Xia (\href{mailto:junxia@hkust-gz.edu.cn}{junxia@hkust-gz.edu.cn})}

\begin{document}
\include{commands.tex}
\maketitle


\input{sec/1_intro}
\input{sec/4_experiments}
\input{sec/5_conclusion}
\input{sec/3_method}

\bibliographystyle{unsrtnat}
\bibliography{ref}

\clearpage
\beginappendix
\input{sec/6_supp}

\end{document}

%% file: commands.tex

\newtheorem{example}{Example}
\newtheorem{definition}{Definition}
\newtheorem{lemma}{Lemma}
\newtheorem{theorem}{Theorem}
\newtheorem{proposition}{Proposition}
\newtheorem{corollary}{Corollary}[proposition]
\newtheorem{assumption}{Assumption}
\newtheorem{observation}{Observation}

\newcommand{\fig}[1]{Fig.~\ref{#1}}
\newcommand{\eq}[1]{Eq.~(\ref{#1})}
\newcommand{\tb}[1]{Tab.~\ref{#1}}
\newcommand{\se}[1]{Section~\ref{#1}}
\newcommand{\ap}[1]{Appendix~\ref{#1}}

\newcommand*{\dif}{\mathop{}\!\mathrm{d}}
\newcommand{\bbE}{\ensuremath{\mathbb{E}}}
\newcommand{\bbR}{\ensuremath{\mathbb{R}}}
\newcommand{\caL}{\ensuremath{\mathcal{L}}}
\newcommand{\caD}{\ensuremath{\mathcal{D}}}


%% file: sec/1_intro.tex
\section{Introduction}
\label{sec:intro}
Small molecules constitute the foundation of chemical diversity, playing a crucial role in 
biological signaling~\cite{fukuto2012small}, metabolic pathways~\cite{qiu2023small} and drug development~\cite{newman2020natural}. In marketed drugs, small molecules account for over 90\% of the total~\cite{southey2023introduction}, demonstrating their clinical significance. To harness the potential of small molecules, the first step is to determine their structures, as their functions are intrinsically linked to their structures. Key features, such as functional groups and chemical bonds, influence the properties of small molecules. However, the chemical space of small molecules is extremely vast, with the majority of it still unexplored. Indeed, it is estimated that the small organic chemical space has more than $10^{60}$ molecules~\cite{kirkpatrick2004chemical, xie2022much}, while PubChem~\cite{kim2025pubchem}, a representative large-scale molecular database, contains on the order of $10^8$ compounds.

\begin{figure}[H]
    \centering
    \includegraphics[width=1\textwidth]{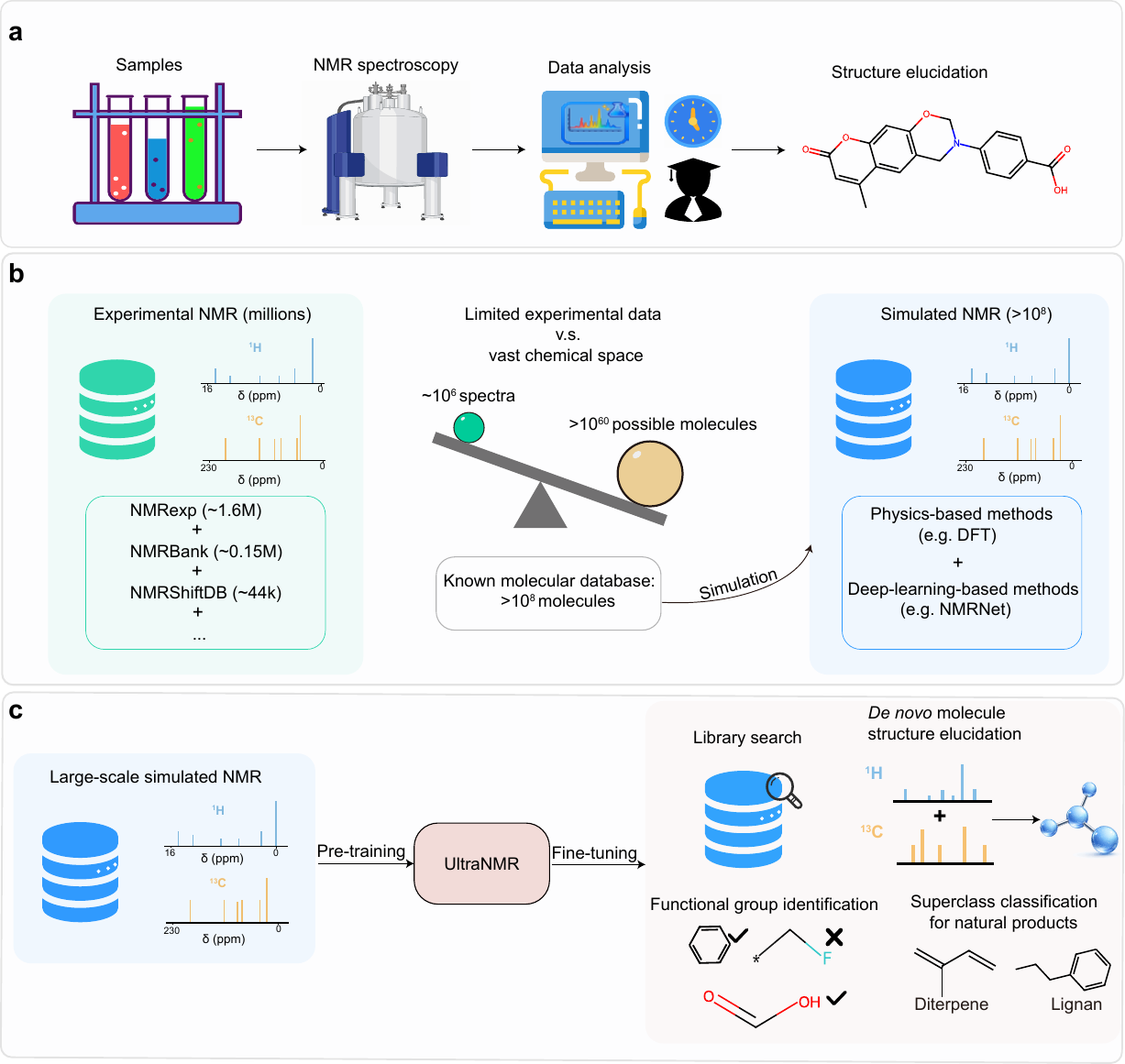}
    \caption{\textbf{Introduction of UltraNMR.} \textbf{a}, Given chemical samples, NMR spectroscopy produces atomic-scale spectral data by probing the magnetic properties of atomic nuclei, which are subsequently analyzed through expert interpretation to determine accurate molecular structures. However, manual interpretation of NMR spectra is highly time-consuming and requires domain expertise. \textbf{b}, Due to the high cost of acquiring NMR spectral data, the largest available experimental NMR databases currently contain only on the order of millions of spectra, which is negligible compared with the vast chemical space (over $10^{60}$) and even known molecular databases (over $10^8$). To bridge this gap, large-scale simulated NMR spectra can be generated from known molecular structures using physics-based methods (e.g., density functional theory (DFT)~\cite{g16}) and deep-learning–based methods (e.g., NMRNet~\cite{xu2025toward}), enabling the construction of expansive simulated NMR datasets for training NMR foundation models. \textbf{c}, We introduce UltraNMR, a foundation model trained on a large-scale simulated NMR dataset, which can be applied to a wide range of NMR interpretation tasks.}
    \label{fig:intro}
\end{figure}

In the realms of biology and chemistry, spectroscopic techniques are essential tools for molecular analysis. Mass spectrometry provides insights into fragmentation patterns and  infrared spectroscopy yields information about functional groups. In contrast, Nuclear Magnetic Resonance (NMR) spectroscopy offers comprehensive structural and dynamical atomic-scale information~\cite{bishop2023robust}, making it a powerful technique for the discovery and identification of small molecules (Fig.~\ref{fig:intro}a). By probing the magnetic properties of atomic nuclei, typically the proton ($^{1}$H) and carbon ($^{13}$C) in a molecule, NMR spectra provide a rich array of parameters, including chemical shifts that reflect local electronic environments and spin-spin coupling constants that reveal the connectivity. However, the interpretation of NMR spectra is time-consuming and requires expert knowledge due to the complexity of NMR spectral data where peak overlap and solvent-induced shifts can obscure critical structural information~\cite{giraudeau2017challenges}. Therefore, automated approaches for NMR spectral analysis are in high demand to accelerate the pace of chemical discovery.

Recently, spectral artificial intelligence approaches have emerged as powerful tools to automate NMR spectral interpretation. By leveraging deep learning algorithms, these methods can rapidly analyze complex NMR spectral data~\cite{10.24963/ijcai.2025/1160,luo2025deep,kim2023deepsat}. Among existing NMR-based molecular analysis tasks, molecular structure elucidation represents a well-studied direction. Current methods can be broadly categorized into library search-based identification and \textit{de novo} elucidation. Library search-based methods identify molecules by comparing query spectra with reference spectra or molecules stored in databases~\cite{jin2025nmr, yang2021cross}. In contrast, \textit{de novo} approaches directly infer molecular structures from NMR spectra without relying on a predefined candidate library~\cite{alberts2023learning, tan2025transformer, xue2025nmrmind, yang2025diffnmr, xiongatomic}. Beyond full structure elucidation, NMR spectra have also been used for other structure-related prediction tasks, including functional group or substructure recognition~\cite{li2022identifying, lee2025machine, hu2024accurate, kuhn2022pilot}, and natural product superclass prediction~\cite{martinez2020prediction}. However, these methods are typically task-specific and lack the generalization needed for broader applications in molecular analysis. As a result, each individual NMR interpretation task requires training a separate model from scratch, limiting their scalability in diverse chemical settings. To address these limitations, foundation models, which learn general representations of data that can be fine-tuned for various downstream applications~\cite{simeoni2025dinov3, zhou2023foundation, bushuiev2025self, cui2024scgpt}, have been proposed as a promising solution.

However, training a foundation model requires large-scale data~\cite{radford2021learning,raffel2020exploring}, which remains scarce in the NMR domain due to the high cost and limited throughput of experimentally acquired spectra. In contrast, simulated NMR spectra can be efficiently generated at scale using simulation techniques  ~\cite{g16,yang2021predicting,gerrard2020impression,xu2025toward}. We therefore collect a large-scale simulated NMR dataset comprising 158 million paired proton ($^{1}$H) and carbon ($^{13}$C) spectra (Fig.~\ref{fig:intro}b), generated via both physics-based~\cite{g16} and deep-learning–based methods~\cite{xu2025toward}. Building upon this dataset, we introduce UltraNMR, an NMR foundation model based on the transformer encoder~\cite{vaswani2017attention}. UltraNMR employs two domain-inspired learning objectives to capture the intrinsic properties of NMR spectra, learning generalizable spectral representations that facilitate seamless simulation-to-real adaptation. The first objective is to predict masked chemical shifts by ordinal regression, enabling UltraNMR to learn the underlying dependencies among chemical shifts within a spectrum. The second objective is inspired by Heteronuclear Single Quantum Coherence (HSQC) spectra, leveraging the known carbon--hydrogen (C--H) bonds to learn the correlation between proton and carbon chemical shifts. Additionally, UltraNMR incorporates molecular fingerprint supervision, aligning spectral embeddings with chemical structures to learn the relationships between spectra based on structural similarity. We also employ contrastive learning on a curated isomer dataset, enabling the model to distinguish structural nuances.

Based on our UltraNMR and simulated NMR spectra dataset, we construct a large-scale NMR spectra library containing 158 million spectra embedding vectors, with around 94 million unique molecules. When conducting library searches on public NMR experimental spectra datasets, we achieve state-of-the-art performance in both accuracy and recall (the proportion of molecules that are actually present in the library and are successfully retrieved), demonstrating the diversity of the data covered by our library and the superiority of the embeddings generated by UltraNMR. We also show that by fine-tuning UltraNMR on diverse experimental NMR-based molecular structure analysis tasks, including \textit{de novo} molecule structure elucidation, functional group prediction and superclass classification of natural products, it consistently achieves better performance compared to other task-specific methods. In real-world applications, we leverage UltraNMR to identify two previously unreported molecules (named Hypoglycin N and Magdiligol G) from \textit{Vitex trifolia} L. var. \textit{simplicifolia} Cham. and \textit{Magnolia officinalis} Rehd. et Wils. var. \textit{biloba} Rehd. et Wils., both of which are valuable Chinese herbal medicines recorded in the Chinese Pharmacopoeia and rich sources of bioactive natural products. 

%% file: sec/4_experiments.tex
\newpage
\section{Results}
\label{sec:experiments}

\subsection{Overview of UltraNMR}
\label{overview}
\begin{figure}[H]
    \centering
    \includegraphics[width=1\textwidth]{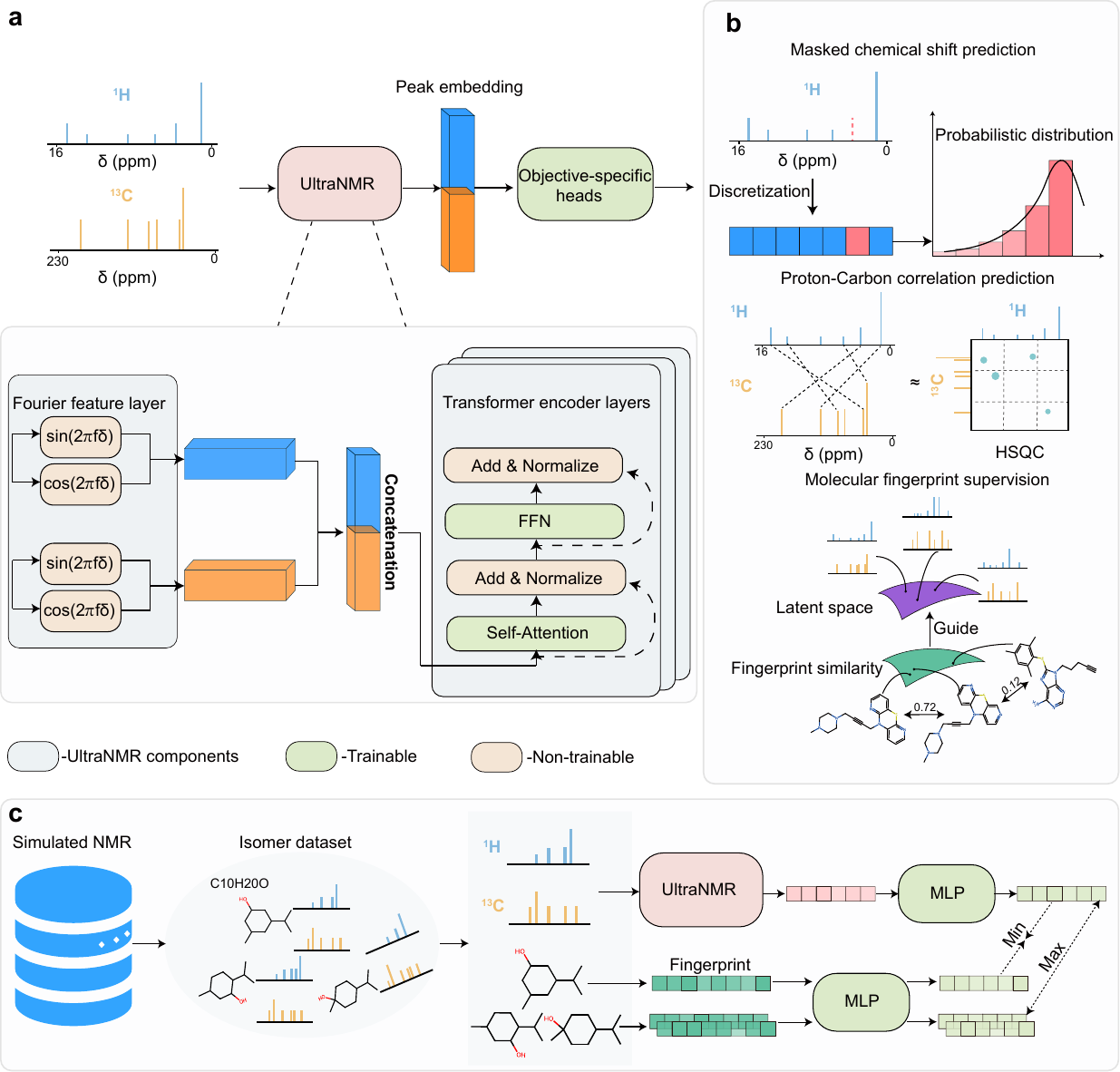}
    \caption{\textbf{Overview of UltraNMR.} \textbf{a}, Model architecture of UltraNMR. Paired $^{1}$H and $^{13}$C chemical shifts are initially encoded via a Fourier feature layer and concatenated. These representations are then processed by transformer encoder layers to generate peak-level spectral embeddings, which are subsequently fed into objective-specific heads for diverse tasks.
\textbf{b}, Pre-training objectives of UltraNMR. The model is trained with masked chemical shift prediction, where masked chemical shifts are recovered from the spectral context via ordinal regression, and proton–carbon correlation prediction, inspired by HSQC spectra to capture inter-atomic dependencies. Additionally, UltraNMR learns to align spectral embeddings with chemical structural similarity through molecular fingerprint supervision. \textbf{c}, Contrastive learning on the curated isomer dataset. To enhance sensitivity to subtle structural differences, we construct a curated isomer dataset by grouping molecules with identical molecular formulas. The model is trained to distinguish the correct molecular fingerprint from those of other isomers via contrastive learning, thereby improving fine-grained structural discrimination.}
    \label{fig:overview}
\end{figure}

UltraNMR is a large transformer-based neural network with 120 million parameters. It consists of 16 standard transformer encoder layers, each composed of multi-head self-attention blocks and feed-forward neural networks~\cite{vaswani2017attention} (Fig.~\ref{fig:overview}a). UltraNMR treats chemical shifts as tokens, encoding them into fixed-dimensional embeddings (768-dimensional vectors), and utilizes mean pooling to aggregate these embeddings into global representations of NMR spectra. To achieve this, paired proton ($^{1}$H) and carbon ($^{13}$C) spectra are first encoded into fixed vectors using a modified Fourier feature approach~\cite{bushuiev2025self}. This method decomposes each chemical shift into predefined sine and cosine frequencies, capturing both integer and fractional components, thereby enhancing the representation of fine-grained spectral details. The processed proton ($^{1}$H) and carbon ($^{13}$C) spectral vectors are then concatenated and fed as input to the transformer model. During pre-training, UltraNMR employs two self-supervised objectives together with molecular fingerprint supervision to learn the intra-spectral and inter-spectral relationships within NMR data (Fig.~\ref{fig:overview}b).

The first objective is masked chemical shift prediction, inspired by the masked language modeling approach used in BERT~\cite{devlin2019bert}. In this task, certain chemical shifts in the NMR spectra are randomly masked, and UltraNMR is trained to predict the masked values. Rather than treating this as a standard regression task, we convert it into an ordinal regression task by discretizing the chemical shift values into bins. We avoid standard regression because it is often overly sensitive to experimental noise and systematic simulation-to-experimental offsets, which can lead to poor generalization. Meanwhile, standard classification is unsuitable for NMR spectra as it treats all bins as independent categories, penalizing a near-miss as severely as a distant category, thus ignoring the inherent physical ordering of the chemical shift scale. By adopting ordinal regression, UltraNMR enforces a distance-aware penalty that aligns with the inherent structure of NMR data, where the proximity between bins reflects the gradual variation in electronic shielding effects and local chemical environments. This enables the model to learn generalizable and physically meaningful representations of NMR spectra. 

The second objective is proton–carbon correlation prediction, inspired by HSQC spectroscopy, which experimentally visualizes the direct linkage between protons and carbons. Since atomic assignments corresponding to chemical shifts are known for simulated spectra, UltraNMR is trained to explicitly model the relationship between proton ($^{1}$H) and carbon ($^{13}$C) chemical shifts that are directly bonded at the molecular level. This enables UltraNMR to encode the precise atom-level connectivity essential for accurate molecule structure elucidation.

To further enhance the model’s perception of structural similarity, we integrate two strategies into the pre-training phase. First, UltraNMR incorporates molecular fingerprint supervision, which aligns the latent spectral embeddings with chemical structures. This supervision allows UltraNMR to learn global molecular topology, ensuring that spectra from structurally similar molecules are mapped closely in the embedding space. Second, we curate a specialized isomer dataset from the training corpus where we train the model to distinguish the structural differences between isomers by contrastive learning, enabling fine-grained structural discrimination (Fig.~\ref{fig:overview}c). 

Large-scale datasets are crucial for foundation models to learn generalizable representations, as they provide a comprehensive range of examples that enable the model to capture underlying relationships in the data~\cite{radford2021learning,raffel2020exploring}. To this end, we construct a comprehensive simulated NMR dataset derived from two distinct sources. The primary subset is the SimNMR-PubChem Database~\cite{jin2025nmr}, comprising 106 million NMR spectra simulated by NMRNet~\cite{xu2025toward}, which is a high-speed deep-learning-based NMR simulation model. To complement this with physics-based simulated data, we collect a secondary subset of 45 million molecules, also sourced from PubChem~\cite{kim2025pubchem} but simulated using Gaussian software~\cite{g16,xue2025nmrmind}. Collectively, the constructed dataset contains 158 million paired proton ($^{1}$H) and carbon ($^{13}$C) spectra, providing a solid foundation for UltraNMR to learn robust and generalizable NMR spectral embeddings.

\subsection{Embedding vector visualization}

To assess whether UltraNMR learns chemically meaningful representations, we visualize the learned global spectral embeddings with uniform
manifold approximation and projection (UMAP) projections~\cite{mcinnes2018umap}. As shown in Fig.~\ref{fig:results1}a, the latent space exhibits a well-organized topology, where structurally related molecules occupy nearby regions, forming a continuous representation of chemical space.

For instance, fatty acids (blue) form a distinct and well-separated cluster on the far right of the projection, and halogenated compounds (pink) are densely concentrated in a separate region on the left. In contrast, carbohydrate-related molecules dominate the central area and display greater dispersion. Notably, carbohydrates containing different heteroatoms (N, S, or P) do not form isolated clusters but instead distribute along smooth manifolds connecting related chemical groups. This continuous transition reflects the hierarchical nature of chemical modifications: as the molecular scaffold shifts from pure carbohydrates to those with amino or phosphate substitutions, their embeddings traverse a smooth trajectory in the latent space. Such connectivity indicates that UltraNMR preserves intrinsic structural similarities.

\subsection{Large-scale NMR library construction}
\label{library search}
Spectral library search serves as a vital approach in computer-assisted structure elucidation (CASE), simplifying the process of molecular identification without the need for complex and time-consuming deduction~\cite{duhrkop2019sirius,bruguiere202113c,yang2021cross}. By matching query spectra against a reference database, we can infer the molecular structures of the query spectra based on the most similar spectra in the library. This approach accelerates the identification process and enhances the accuracy of structure elucidation, particularly for molecules with known chemical shifts. However, the success of spectral library search is heavily dependent on the quality and comprehensiveness of the library itself. Due to the high cost and time required to acquire experimental NMR data, the size of established NMR libraries remains limited. Up to now, the largest experimental NMR database contains only 3.3 million entries~\cite{wang2025nmrexp}, representing only a small fraction of the entire chemical space. In contrast, simulated NMR libraries can be generated at scale using physics-based or deep-learning-based methods~\cite{g16,xu2025toward}. However, traditional spectral library search methods, such as peak-to-peak matching, primarily retrieve spectra with high signal-level similarity, which does not necessarily guarantee structural similarity between the corresponding molecules. UltraNMR addresses this challenge by learning generalizable spectral embeddings, capturing both intra- and inter-spectral relationships as well as the chemical significance underlying the spectra. By forwarding our constructed NMR dataset into UltraNMR, we build a large-scale simulated NMR library that contains 158 million spectral vectors, with around 94 million unique molecules. In this library, each NMR spectrum is encoded as a 768-dimensional vector.

We evaluate our constructed NMR library on NMRGym~\cite{fang2026nmrgym}, the largest experimental NMR spectra benchmark to date. We compare embeddings produced by UltraNMR with those obtained using the Gaussian convolution method adopted in NMR-Solver, a traditional method that converts an NMR spectrum into a fixed-dimensional vector. During library search, a query spectrum consisting of paired proton ($^{1}$H) and carbon ($^{13}$C) NMR signals is first encoded into a fixed-dimensional vector using either UltraNMR or Gaussian convolution. We then retrieve candidate spectra whose molecules share the same molecular formula as the query. We compute the cosine similarity between the query embedding and each candidate embedding to identify the most similar spectrum. The molecule corresponding to the top-ranked spectrum (i.e., with the highest cosine similarity) is reported as the predicted structure for the query spectrum. 

Fig.~\ref{fig:results1}b summarizes the overall library search performance on NMRGym. Across different retrieval depths, UltraNMR consistently outperforms Gaussian convolution in both accuracy and recall. We also compare UltraNMR with NMRSolver on a subset of the NMRGym test set. As shown in Fig.~\ref{fig:results1}c, UltraNMR consistently achieves higher retrieval accuracy than NMRSolver under both stereochemistry-aware and stereochemistry-agnostic evaluation settings. To examine the structural relevance of the retrieved molecules, we analyze the Tanimoto similarity distribution of Top-1, Top-5, and Top-10 predictions (Fig.~\ref{fig:results1}d). Compared with Gaussian convolution, UltraNMR yields a higher proportion of predictions in the high-similarity regime, while Gaussian convolution exhibits a larger fraction of low-Tanimoto matches. This shift in distribution suggests that UltraNMR is more likely to retrieve molecules that are structurally close to the ground truth. These results demonstrate the superiority of UltraNMR embeddings and the broad chemical diversity covered by our large-scale spectral library.

We further dissect the failure cases by comparing the Top-1 errors of the two methods across different Tanimoto similarity intervals (Fig.~\ref{fig:results1}e). For the same set of query spectra, cases where UltraNMR retrieves a high-similarity molecule while Gaussian convolution retrieves a low-similarity one are more frequent than the reverse scenario. For example, UltraNMR achieves a Tanimoto similarity of 0.6–0.8 while Gaussian convolution yields 0.0–0.2 in 235 cases, whereas the reverse occurs in only 139 cases. This asymmetry indicates that scenarios in which traditional signal-level matching fails but UltraNMR remains effective are more common than the opposite case.
\vspace{-1.8em}
\begin{figure}[H]
    \centering
    \includegraphics[width=1\textwidth]{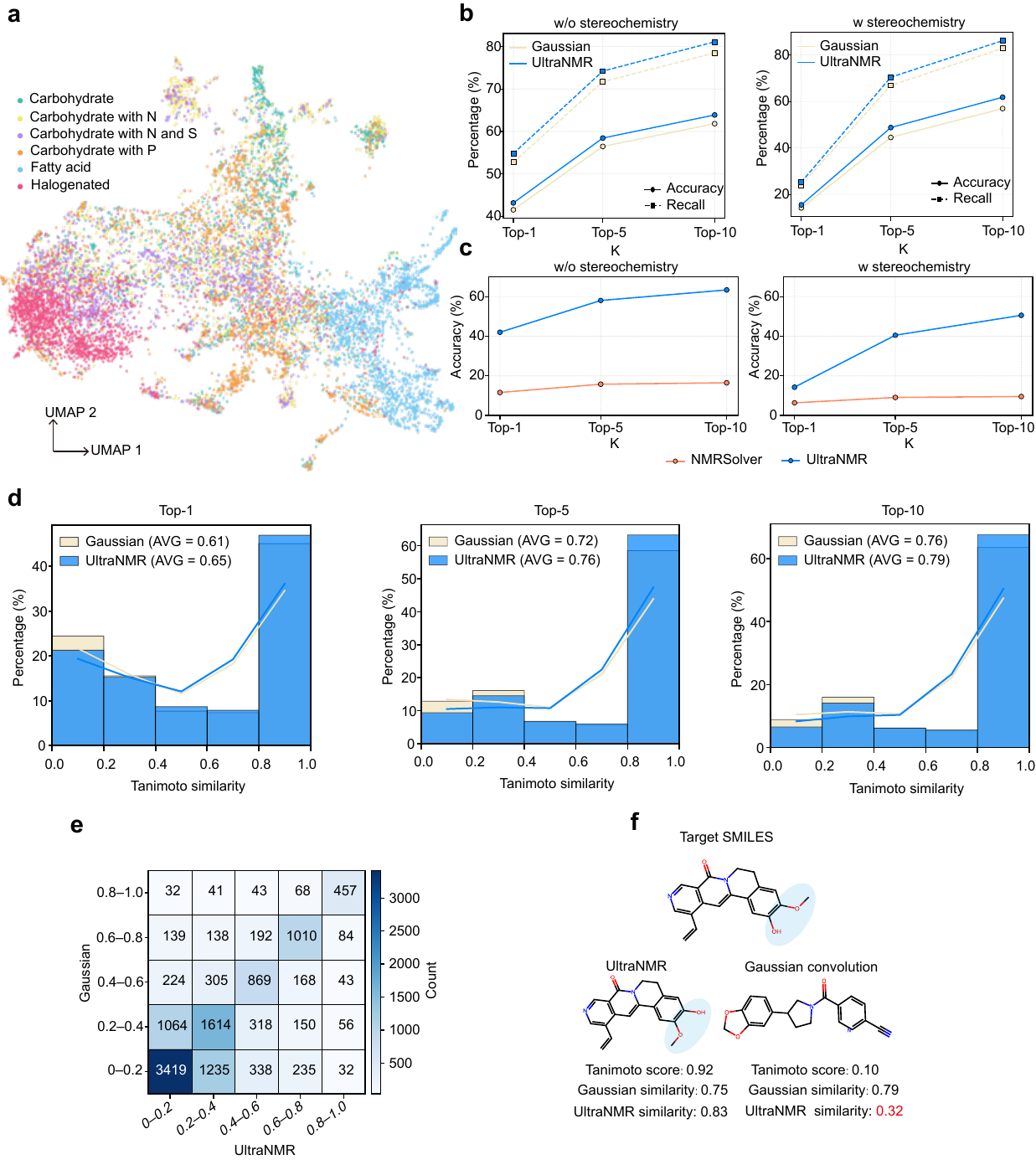}
\end{figure}

\begin{figure}[H]
    \ContinuedFloat
    \renewcommand{\thefigure}{3}
    \caption{\textbf{Direct evaluation of UltraNMR embeddings.}
\textbf{a}, UMAP visualization of global spectral embeddings learned by UltraNMR, with points colored by molecular classes. Fatty acids, halogenated compounds and pure carbohydrates are clearly clustered. \textbf{b}, Overall library search performance with and without stereochemistry across different retrieval depths on the NMRGym benchmark. UltraNMR consistently outperforms Gaussian convolution in both accuracy and recall. 
\textbf{c}, Comparison between UltraNMR and NMRSolver on a subset of the NMRGym test set. UltraNMR achieves consistently higher retrieval accuracy across different retrieval depths.
\textbf{d}, Distribution of Tanimoto similarity between Top-1, Top-5, and Top-10 retrieved molecules and target structures. Compared with Gaussian convolution, UltraNMR retrieves structurally closer molecules with consistently higher proportions in similarity ranges above 0.5. \textbf{e}, Comparison of Top-1 error cases across Tanimoto similarity intervals for UltraNMR and Gaussian convolution. UltraNMR more frequently retrieves high-similarity candidates in cases where Gaussian convolution produces low-similarity errors. \textbf{f}, Representative library search example where both methods fail to retrieve the exact target at Top-1. This case shows the limitation of signal-level similarity and the advantage of UltraNMR embeddings in capturing structure-level chemical relationships.}
    \label{fig:results1}
\end{figure}
\vspace{-1em}
A representative example is shown in Fig.~\ref{fig:results1}f. For this query spectrum, both methods fail to retrieve the exact target molecule at Top-1. However, UltraNMR retrieves a candidate with high structural similarity to the target (Tanimoto score = 0.83), whereas Gaussian convolution retrieves a molecule with much lower structural similarity (Tanimoto score = 0.10), despite exhibiting a high signal-level spectral similarity (Gaussian similarity = 0.79). Notably, the molecule retrieved by Gaussian convolution also shows low similarity in the UltraNMR embedding space (UltraNMR similarity = 0.32). This example highlights a fundamental limitation of signal-level similarity metrics and illustrates that UltraNMR embeddings can capture chemically meaningful relationships.

\subsection{Transfer learning to NMR interpretation tasks}
\label{downstream}
As a foundation model, UltraNMR learns generalizable NMR spectral embeddings that can be effectively transferred to various downstream spectral interpretation tasks. We evaluate UltraNMR on three structure-centric tasks: superclass classification of natural products, functional group identification and \textit{de novo} molecule structure elucidation. Following the library search setting, we use the NMRGym benchmark~\cite{fang2026nmrgym} to compare UltraNMR with state-of-the-art methods. We also include an UltraNMR variant trained from scratch (without pre-training) as a baseline to evaluate the effectiveness of pre-training. For each task, we attach a task-specific decoder to UltraNMR and fine-tune the entire model. For the functional group identification task and the superclass classification task, we employ a simple linear head, which is commonly used in transfer learning~\cite{chen2020simple, assran2023self, he2020momentum}. For \textit{de novo} structure elucidation, we adopt a transformer-based decoder. To prevent data leakage, we excluded any molecules present in the downstream test sets from the simulated training data. All results are reported as the average of three independent runs with different random seeds.

For the functional group identification task, each spectrum is labeled with a 20-dimensional binary vector where each dimension indicates the presence (1) or absence (0) of a specific functional group. As shown in Fig.~\ref{fig:ds}a and b, UltraNMR consistently outperforms the state-of-the-art methods across all evaluation metrics. Specifically, it reaches an overall accuracy of 50.20\%, surpassing the previous best method (45.68\% by XGBoost). UltraNMR also shows clear advantages on macro-averaged scores. It improves macro-recall by about 5\% over the previous best method (NMR2Struct~\cite{hu2024accurate}), and increases macro-F1 to 65.14\%, higher than the strongest baseline (55.47\% by XGBoost). These gains indicate that UltraNMR is not only strong on frequent functional groups, but also more reliable on rare and hard-to-predict groups. To further validate the representational power of the learned embeddings, we also compare UltraNMR embeddings with Gaussian-vectorized spectra used in Section~\ref{library search} as inputs for the task-specific decoder. The results show that UltraNMR outperforms the Gaussian vector with a 10\% higher accuracy, which demonstrates that the spectral embeddings generated by UltraNMR capture high-level chemical semantics rather than just raw spectral patterns.

For the superclass classification task, we use NPClassifier~\cite{kim2021npclassifier} to label NMR spectra. NPClassifier is a deep-learning-based tool specifically designed for the automated structural classification of natural products. In our study, spectra are categorized into 74 distinct superclasses based on their corresponding molecules. We filter out the spectra whose corresponding molecules cannot be classified by NPClassifier. As shown in Fig.~\ref{fig:ds}c, UltraNMR shows improved performance over the baseline model across all common superclasses (those with more than 500 test samples). We also highlight the superclasses exhibiting the largest relative performance gains (greater than 20\% improvement with more than 100 test samples). Notably, these classes predominantly correspond to aromatic compounds. This trend suggests that UltraNMR is particularly effective at capturing characteristic spectral patterns associated with aromatic ring systems and conjugated structures.

\setcounter{figure}{3}  
\begin{figure}[H]
    \centering
    \includegraphics[width=1\textwidth]{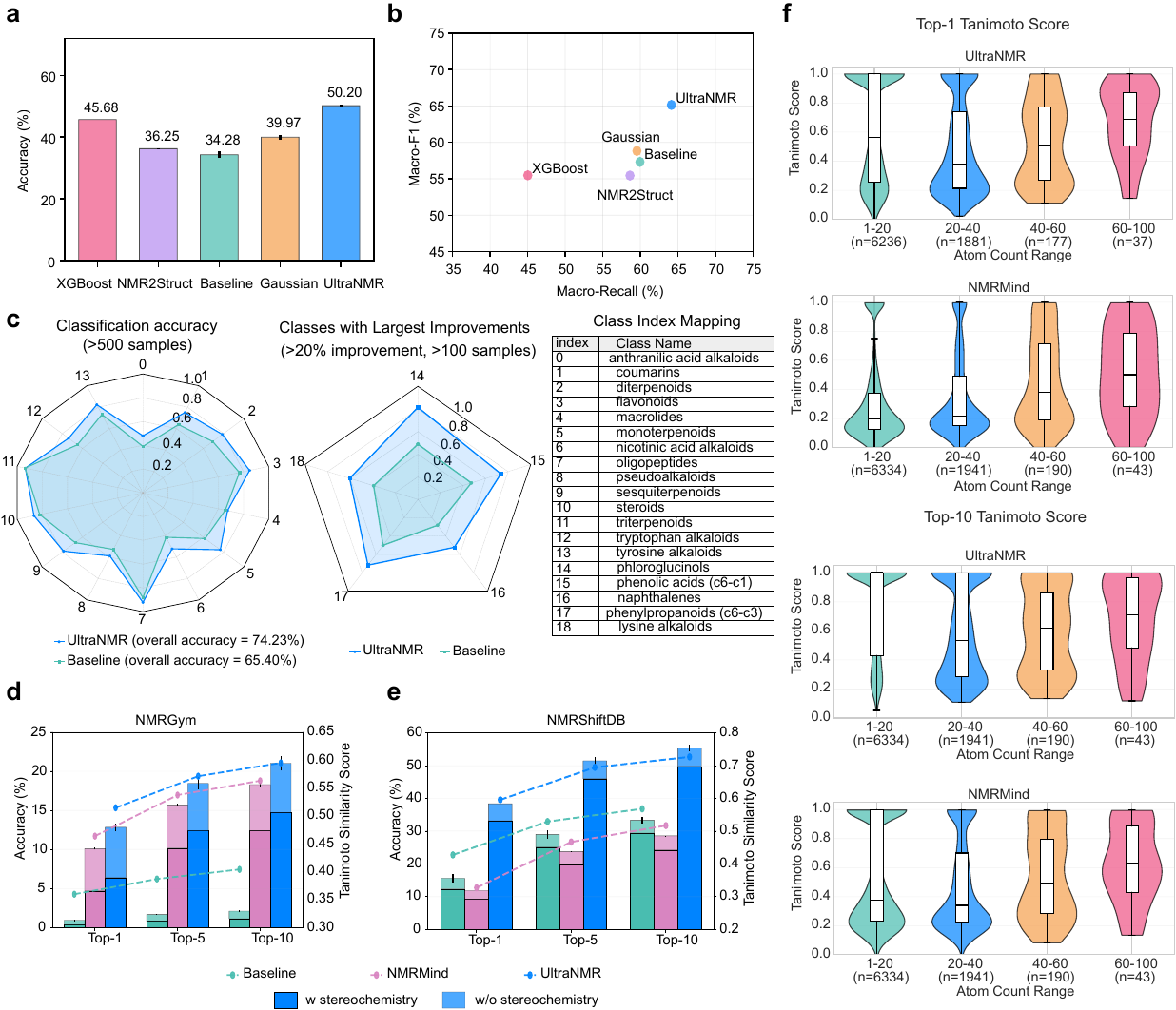}
        \caption{\textbf{Evaluation of UltraNMR on downstream tasks}. \textbf{a}, Functional group identification accuracy on the NMRGym benchmark. The results of XGBoost and NMR2Struct are from NMRGym~\cite{fang2026nmrgym}. UltraNMR outperforms both traditional machine learning (XGBoost) and deep learning baselines (NMR2Struct). \textbf{b}, Comparison of Macro-F1 and Macro-Recall for functional group identification. UltraNMR occupies the top-right corner, indicating superior performance on both frequent and rare functional groups. \textbf{c}, Performance on superclass classification. Radar plots (left and center) show classification accuracy for common superclasses and those with the largest improvements compared to the baseline model. The index mapping (right) details the specific natural product classes. \textbf{d, e}, \textit{De novo} molecule structure elucidation performance on NMRGym (d) and NMRShiftDB (e) datasets. Accuracy for Top-1, Top-5, and Top-10 predictions is shown for structures with (darker bars) and without (lighter bars) stereochemistry. The dashed lines represent the average Tanimoto similarity scores. \textbf{f}, Structural similarity distribution across different molecule sizes on the NMRShiftDB dataset. UltraNMR shows higher median scores and a greater density of high-similarity structures compared to NMRMind. UltraNMR consistently achieves the highest accuracy and structural similarity, demonstrating the robust transferability of its pre-trained spectral embeddings to diverse chemical environments.}
    \label{fig:ds}
\end{figure}

For the \textit{de novo} molecule structure elucidation task, we generate SMILES from NMR spectra with molecular formulas using UltraNMR with attached transformer decoder layers. To fully leverage the simulated spectra data, we employ a two-stage training strategy: large-scale sequence-to-sequence training on simulated spectra to learn how to map spectra to SMILES, followed by fine-tuning on downstream datasets. Since the NMRGym dataset utilizes a scaffold-based split, the training and test sets represent distinct distributions. Under this challenging setting, the baseline model performs poorly, while pre-trained models, i.e., NMRMind and UltraNMR, show significant improvements. As shown in Fig.~\ref{fig:ds}d, UltraNMR consistently outperforms NMRMind across all metrics, with about 2\% improvement. Compared to the baseline model, UltraNMR achieves a nearly 40-fold increase in Top-1 accuracy for structures including stereochemistry and higher Tanimoto similarity scores, indicating the necessity of pre-training. On the NMRShiftDB dataset~\cite{kuhn2015facilitating,xu2025toward}, we adopt a random split, which results in a less challenging evaluation setting compared to the scaffold-based split used in NMRGym. 
As shown in Fig.~\ref{fig:ds}e, in this setting, the baseline model already achieves strong performance and even surpasses the pre-trained NMRMind model, which highlights the superiority of our model architecture. However, even with this strong baseline, incorporating large-scale pre-training and sequence-to-sequence training leads to large improvements. UltraNMR more than doubles the Top-1 accuracy of the baseline model for both stereochemical and non-stereochemical tasks. In addition, the Tanimoto similarity scores increase noticeably, demonstrating that the pre-trained UltraNMR is not only more accurate in molecule structure elucidation but also exhibits a deeper understanding of chemical space. Furthermore, we compare the performance of UltraNMR with NMRMind on the NMRShiftDB dataset using Tanimoto scores across different molecular sizes (Fig.~\ref{fig:ds}f). For both Top-1 and Top-10 metrics, UltraNMR consistently achieves higher median Tanimoto scores and more concentrated distributions in the high-similarity region compared to NMRMind. This advantage is particularly pronounced in smaller molecules (1-20 atoms) and remains robust as molecular complexity increases (60-100 atoms).

\subsection{Structure elucidation of novel natural products}

Natural products often exhibit highly complex and diverse molecular structures, which makes their structural elucidation particularly challenging. In practice, resolving such structures typically requires expert-level knowledge and extensive manual analysis. UltraNMR can simultaneously perform superclass classification, functional group identification, spectral library search, and \textit{de novo} structure elucidation. By integrating the complementary information, UltraNMR allows experts to focus on targeted deduction rather than exhaustive exploration.

By leveraging the predictions of UltraNMR, we identify two previously unreported molecules from \textit{Vitex trifolia} and \textit{Magnolia officinalis}, both of which are valuable traditional herbal medicines. In both cases, we first generate a candidate pool under molecular formula constraints using spectral library search and \textit{de novo} molecule structure elucidation. We then refine this candidate set based on the superclass and functional group predictions provided by UltraNMR and submit the filtered candidates for expert assessment.

For the first molecule from \textit{Vitex trifolia}, UltraNMR correctly predicts the diterpene superclass, with ketone and alkene groups. After applying these filters, the initially top-ranked \textit{de novo} candidate is excluded because its molecular formula is inconsistent with the given formula. Consequently, the originally second-ranked \textit{de novo} structure becomes the top-ranked candidate and is confirmed as the correct molecule. In contrast, the best candidates retrieved via library search all exhibit low structural similarity to the ground truth, with Tanimoto scores below 0.2, indicating the absence of close references in the database. In this scenario, \textit{de novo} prediction plays a critical role in elucidating the correct structure.

For the second molecule from \textit{Magnolia officinalis}, UltraNMR does not directly recover the exact structure, likely due to its higher structural complexity compared to the first case. However, UltraNMR correctly predicts the lignan superclass, along with functional groups including aromatic rings, alcohols, alkenes, and ethers. Applying these constraints to filter the candidate pool reduces the search space. The top two candidates from library search after filtering exhibit high structural similarity to the target molecule, with Tanimoto scores of 0.62. These candidates differ from the true structure only in minor structural variations, specifically the size of the oxygen heterocycle. These close matches provide valuable structural cues, effectively guiding experts toward the correct structure.

\begin{figure}[H]
    \centering
    \includegraphics[width=1\textwidth]{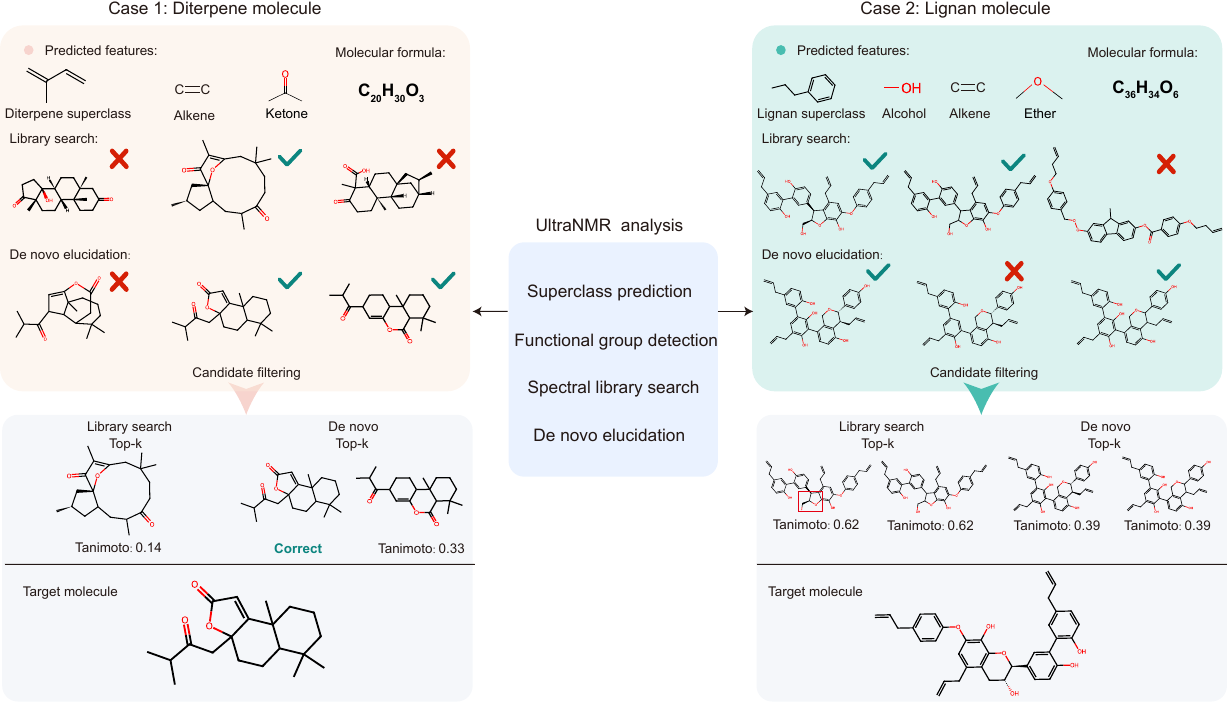}
    \caption{\textbf{Structure elucidation of two novel natural products.} 
    \textbf{Case 1}, Identification of a novel diterpene molecule from \textit{Vitex trifolia}. UltraNMR correctly identifies the diterpene superclass and functional groups (ketone and alkene). Despite the failure of library search to retrieve close references (Tanimoto score < 0.20), the \textit{de novo} structure elucidation model successfully generates the correct structure within the top three predictions.
    \textbf{Case 2}, Identification of a novel lignan molecule from \textit{Magnolia officinalis}. UltraNMR correctly predicts the lignan superclass and functional groups (alcohol, alkene, and ether). Although UltraNMR does not directly yield the exact structure, the top-ranked candidates from library search exhibit high similarity to the target, with the primary difference being the heterocycle ring size (highlighted in red box).
    These cases suggest the potential of UltraNMR to integrate multi-task predictions for the identification of novel natural products.}
    \label{fig:real-world}
\end{figure}

%% file: sec/5_conclusion.tex
\section{Discussion}
\label{sec:conclusion}

In this work, we introduce UltraNMR, a foundation model specifically designed for Nuclear Magnetic Resonance (NMR) spectroscopy. By leveraging a large-scale simulated dataset comprising 158 million paired proton ($^{1}$H) and carbon ($^{13}$C) spectra, UltraNMR learns generalizable spectral representations through a combination of self-supervised objectives and molecular fingerprint supervision. These objectives enable the model to capture intra-spectral dependencies and inter-atomic relationships while aligning spectral embeddings with chemical structures. Extensive experiments demonstrate that the representations learned by UltraNMR are both chemically meaningful and highly generalizable. In large-scale spectral library search, UltraNMR outperforms traditional signal-level similarity methods, retrieving molecules that are structurally closer to the ground truth. Furthermore, when fine-tuned on diverse downstream tasks, UltraNMR consistently achieves state-of-the-art performance on experimental NMR benchmarks. In real-world applications, UltraNMR facilitates the identification of two novel molecules from valuable Chinese herbal medicines recorded in the Chinese Pharmacopoeia, demonstrating its potential to accelerate the identification of novel compounds from natural sources. These results highlight the practical utility of the spectral representations learned by UltraNMR.

Although UltraNMR has demonstrated strong performance across diverse experimental NMR-based molecular structure analysis tasks, several directions remain for further improvement. First, UltraNMR currently focuses on one-dimensional $^{1}$H and $^{13}$C spectra, whereas practical structure elucidation often benefits from complementary NMR information such as HSQC, HMBC, COSY, NOESY, multiplicity, and coupling constants. Extending UltraNMR to integrate multi-dimensional and multi-modal spectroscopic data can improve its ability to resolve challenging isomers, stereochemical configurations, and highly complex natural products. Second, although UltraNMR improves spectral library search and \textit{de novo} structure elucidation performance, exact structure determination remains difficult for molecules with rare scaffolds, subtle structural differences, or limited representation in the training and retrieval libraries. Future work could combine UltraNMR with uncertainty estimation and iterative candidate generation to further improve reliability. These directions may enable UltraNMR to evolve from a powerful representation model into a more comprehensive and practical platform for automated molecular structure analysis.

%% file: sec/3_method.tex
\section{Method}
\label{sec:method}

\subsection{Model architecture}
\label{sec:architecture}

UltraNMR consists of two main modules: a Fourier feature layer and transformer encoder layers, and produces diverse outputs via objective-specific heads. The Fourier feature layer maps input proton and carbon chemical shifts ($\delta_H$ and $\delta_C$) into high-dimensional vectors, providing a compatible representation for the subsequent transformer architecture. The transformer encoder layers then employ self-attention mechanisms to capture the dependencies inherent in NMR spectra. Finally, the objective-specific heads map the learned representations to the curated pre-training objectives, enabling the model to comprehensively learn both intra- and inter-spectral relationships while preserving the underlying chemical significance of the spectra.

Since NMR chemical shifts are continuous, they are not suitable as direct input to a transformer model, which has difficulty learning high-frequency functions~\cite{rahaman2019spectral, tancik2020fourier}. To overcome this limitation and capture fine-grained spectral details, we process each chemical shift using a Fourier feature mapping $\Phi: \mathbb{R} \rightarrow [-1, 1]^{2F}$, parameterized by a vector of $F$ predefined frequencies $\mathbf{F} \in \mathbb{R}^F$. Specifically, the features are constructed using sine and cosine functions:

\begin{equation}
\Phi(\delta)_{i} = \sin(2\pi f_i \delta), \quad \Phi(\delta)_{i+1} = \cos(2\pi f_i \delta),
\end{equation}

where each frequency $f_i$ corresponds to a specific scale of the spectral domain, forming a vector of frequencies that encodes both integer-level and decimal-level information of chemical shifts:

\begin{equation}
\mathbf{f} =
\left[
\frac{1}{\delta_{\max}},
\frac{1}{\delta_{\max}-1},
\ldots,
1,
\;
\frac{1}{k \delta_{\min}},
\frac{1}{(k-1)\delta_{\min}},
\ldots,
\frac{1}{\delta_{\min}}
\right]^{\top}
\in \mathbb{R}^{F},
\end{equation}

where $\delta_{\text{max}}$ denotes the maximum chemical shift range of interest (e.g., the spectral width), 
$\delta_{\text{min}}$ corresponds to the minimum distinguishable shift difference, 
and $k \in \mathbb{N}$ is chosen such that $k\,\delta_{\text{min}}$ is closest to $1$. The Fourier features are then projected to the transformer encoder space via a feed-forward neural network: $\mathbb{R}^{2F} \rightarrow \mathbb{R}^{d_m}$, where $d_m$ denotes the transformer encoder dimension. The proton and carbon chemical shifts are separately encoded via the Fourier feature layer and feed-forward neural networks (FFNs), and the resulting representations are stacked as the input to the transformer encoder:
\begin{equation}
\mathbf{x}
=
\big[
\mathrm{FFN}_{H}(\Phi(\delta_{H})),
\;
\mathrm{FFN}_{C}(\Phi(\delta_{C}))
\big],
\end{equation}
where $\mathbf{x} \in \mathbb{R}^{n \times d_{\text{model}}}$, and
$n$ denotes the total number of chemical shifts from both proton and carbon spectra. $\mathrm{FFN}_{H}$ and $\mathrm{FFN}_{C}$ are modality-specific FFNs for proton and carbon chemical shifts, respectively. Notably, we do not encode positional information into the input $\mathbf{x}$, as chemical shifts form a set rather than a sequence.

Furthermore, the stacked representation $\mathbf{x}$ is processed by a series of transformer encoder layers. Each layer employs a Multi-Head Self-Attention (MSA)~\cite{vaswani2017attention} mechanism followed by an FFN, enabling information exchange between chemical shifts in the spectra. Specifically, for each layer $l \in \{1, \dots, L\}$, the hidden state $\mathbf{h}^{(l)}$ is updated via residual connections~\cite{he2016deep} and layer normalization (LN)~\cite{ba2016layer} as follows:

\begin{align}
\mathbf{z}^{(l)} &= \mathrm{LN}\big(\mathbf{h}^{(l-1)} + \mathrm{MSA}(\mathbf{h}^{(l-1)})\big), \\
\mathbf{h}^{(l)} &= \mathrm{LN}\big(\mathbf{z}^{(l)} + \mathrm{FFN}(\mathbf{z}^{(l)})\big),
\end{align}
where $\mathbf{h}^{(l)} \in \mathbb{R}^{n \times d_{\text{model}}}$ represents the hidden representations at layer $l$, and $\mathbf{h}^{(0)} = \mathbf{x}$. We replace the ReLU activation in the original transformer with the GELU activation function~\cite{hendrycks2016gaussian} and obtain global spectral representations $\tilde{\mathbf{h}}$ via mean pooling over the final layer output $\mathbf{h}^{(L)}$.

We then append objective-specific heads to the transformer encoder, projecting $\mathbf{h}^{(L)}$ or $\tilde{\mathbf{h}}$ to different prediction spaces tailored for each pre-training objective (see details in Section \ref{sec:training}).

\subsection{Pre-training Objectives}
\label{sec:training}
UltraNMR is pre-trained in two stages. In the first stage, the model is trained on the full simulated NMR data using two self-supervised objectives together with molecular fingerprint supervision to learn generalizable representations of NMR spectra. In the second stage, we further refine the learned representations using a constructed isomer dataset, where we introduce an additional isomer contrastive learning objective to facilitate fine-grained spectral representation learning.

\textbf{Masked chemical shift prediction.}
The first self-supervised objective is masked chemical shift prediction, which aims to learn intra-spectral representations by predicting missing chemical shifts from their spectral context. For each proton ($^{1}$H) or carbon ($^{13}$C) spectrum, we randomly sample a subset of chemical shifts and replace their original values with masked placeholders.

During the pre-training stage, only chemical shifts are encoded, without any auxiliary information. If all masked shifts are replaced by a single constant value, the resulting embeddings become identical, causing the prediction task to fail. To address this issue, we introduce an order-aware masking strategy that preserves the ordinal structure among masked shifts without information leakage. We first sort the masked shifts in ascending order according to their original chemical shift values:
\begin{equation}
\delta_{(1)} \le \delta_{(2)} \le \dots \le \delta_{(|\mathcal{M}|)},
\end{equation}
where $\mathcal{M}$ denotes the subset selected for masking, and $|\mathcal{M}|$ is the total number of masked shifts. We then replace each masked shift $\delta_{(k)}$ with a unique negative placeholder value:
\begin{equation}
\tilde{\delta}_{(k)} = -k,\quad k=1,2,\dots,|\mathcal{M}|.
\end{equation}
This strategy ensures that each masked peak receives a distinct embedding. 

The prediction of masked chemical shifts is formulated as an ordinal regression problem rather than standard regression or classification. We first discretize the continuous chemical shift range $[0, \delta_{\max}]$ into $B$ ordered bins with a bin size of $\Delta$, where $B = \left\lfloor \delta_{\max}/\Delta \right\rfloor + 1$. The center of the $j$-th bin is defined as $c_j = j \cdot \Delta$ for $j \in \{0, \dots, B-1\}$.

Instead of assigning a single hard target bin, we construct Gaussian soft labels, which are commonly adopted in ordinal regression tasks~\cite{wang2025survey,tan2016age}, to encode the ordinal structure of chemical shifts. Given a ground-truth chemical shift $\delta_i$ where $i \in \mathcal{M}$, the target probability distribution $\tilde{\mathbf{y}}_i$ over the bins is computed based on the distance between $\delta_i$ and the bin centers. The probability for the $j$-th bin is given by:

\begin{equation}
\label{eq:soft_label}
\tilde{y}_{i,j} = \dfrac{\exp\!\left( - \dfrac{(\delta_i - c_j)^2}{2\sigma^2} \right)}{\sum_{k=0}^{B-1} \exp\!\left( - \dfrac{(\delta_i - c_k)^2}{2\sigma^2} \right)},
\end{equation}

where $\sigma$ is the standard deviation of the Gaussian kernel. Specific values of $\Delta$, $\delta_{\max}$, and $\sigma$ are employed for $^{1}$H and $^{13}$C spectra respectively.

To obtain the predicted probability distribution, the final hidden representation $\mathbf{h}_i^{(L)} \in \mathbb{R}^{d_{\text{model}}}$ corresponding to the $i$-th masked shift is projected to the discretized output space via a linear prediction head:

\begin{equation}
\label{eq:linear_head}
\mathbf{z}_i = \mathbf{W}_{\text{mask}} \mathbf{h}_i^{(L)} + \mathbf{b}_{\text{mask}},
\end{equation}

where $\mathbf{W}_{\text{mask}} \in \mathbb{R}^{B \times d_{\text{model}}}$ and $\mathbf{b}_{\text{mask}} \in \mathbb{R}^{B}$ are learnable weights and biases, and $\mathbf{z}_i \in \mathbb{R}^{B}$ denotes the predicted logits.

The final ordinal loss function is defined as:

\begin{equation}
\label{eq:ord_loss}
\mathcal{L}_{\text{mask}} = - \frac{1}{|\mathcal{M}|} \sum_{i \in \mathcal{M}} \sum_{j=0}^{B-1} \tilde{y}_{i,j} \log \left( \frac{\exp(z_{i,j})}{\sum_{k=0}^{B-1} \exp(z_{i,k})} \right),
\end{equation}

where $z_{i,j}$ denotes the $j$-th element of the logit vector $\mathbf{z}_i$. This formulation effectively penalizes predictions based on their distance from the ground truth, ensuring that the model learns the continuous nature of chemical shifts within the discretized output space and remains robust to systematic simulation-to-experimental offsets.

\textbf{Proton–carbon correlation prediction.}
The second self-supervised objective is proton–carbon correlation prediction. Since we pre-train UltraNMR on simulated spectra with known atom assignments, we supervise the model to align each proton shift with the carbon shift that is directly bonded at the molecular level.

Let $\{\mathbf{h}^{H}_{i}\}_{i=1}^{N_H}$ and $\{\mathbf{h}^{C}_{j}\}_{j=1}^{N_C}$ denote the chemical shift embeddings from the final transformer encoder layer in paired proton and carbon spectra, respectively (the layer index is omitted for simplicity). For a specific proton shift $i$, $j(i)$ denotes the index of its directly bonded carbon shift.

To encourage the model to capture atom-level connectivity across paired spectra, we adopt an InfoNCE-style contrastive objective. We first compute the cosine similarity between the embeddings of proton $i$ and any carbon $j$:

\begin{equation}
\mathrm{sim}_{i,j}
=
\frac{
(\mathbf{h}^{\text{H}}_{i})^{\top}\mathbf{h}^{\text{C}}_{j}
}{
\|\mathbf{h}^{\text{H}}_{i}\|_2 \, \|\mathbf{h}^{\text{C}}_{j}\|_2
}.
\end{equation}

The overall proton–carbon correlation loss is calculated as:

\begin{equation}
\mathcal{L}_{\mathrm{HC}}
=
- \frac{1}{|\mathcal{I}|}
\sum_{i \in \mathcal{I}}
\log
\frac{
\exp\!\left(\mathrm{sim}_{i, j(i)} / \tau \right)
}{
\sum_{k=1}^{N_C} \exp\!\left(\mathrm{sim}_{i, k} / \tau \right)
},
\end{equation}

where $\mathcal{I}$ denotes the set of proton shifts with known bonded carbon counterparts, and $\tau$ is a temperature hyperparameter set to 0.07. 
The model is optimized to maximize the similarity between the bonded pair $(i, j(i))$ while minimizing the similarity with all other non-bonded carbon shifts within the paired spectra. As a result, the learned representations capture atom-level connectivity, which is important for molecule structure elucidation.

\textbf{Molecular fingerprint supervision.}
To further enhance the model's perception of global structural similarity, we incorporate molecular fingerprint supervision during pre-training, which encourages the learned spectral representations to preserve molecular topology, such that spectra from structurally similar molecules are mapped closer in the representation space. We first convert SMILES into Morgan fingerprints, which are fixed-length binary vectors (length 2048, radius 2). Then we calculate the Tanimoto similarity $t_{mn}\in[0,1]$ based on pairwise fingerprints $\mathbf{p}_m$ and $\mathbf{p}_n$ as a structure-level supervisory signal between spectral embeddings:
\begin{equation}
t_{mn}
=
\frac{
\langle \mathbf{p}_m, \mathbf{p}_n \rangle
}{
\|\mathbf{p}_m\|_1 + \|\mathbf{p}_n\|_1 - \langle \mathbf{p}_m, \mathbf{p}_n \rangle
},
\label{eq:tanimoto}
\end{equation}
where $\langle \mathbf{p}_m, \mathbf{p}_n \rangle$ counts the number of shared active bits (intersection), and $\|\cdot\|_1$ counts the number of active bits.

We formulate fingerprint supervision as a classification task with focal loss~\cite{lin2017focal} rather than a regression task~\cite{xiong2025supervised}, which enables the model to better handle the highly imbalanced distribution of molecular similarity pairs. We first discretize $t_{mn}$ into ordered bins with the bin size $\Delta_{\text{fp}}$:
\begin{equation}
y_{mn}
=
\left\lfloor \frac{t_{mn}}{\Delta_{\text{fp}}} \right\rfloor,
\qquad
y_{mn} \in \{0,1,\dots,K-1\},
\label{eq:tanimoto_bin}
\end{equation}
where $K = \left\lfloor 1/\Delta_{\text{fp}} \right\rfloor + 1$. We then concatenate the pairwise global spectral representations $(\tilde{\mathbf{h}}_m,\tilde{\mathbf{h}}_n)$ and predict the similarity bin using a multilayer perceptron (MLP) classifier $g(\cdot)$ followed by a softmax function:
\begin{equation}
\mathbf{o}_{mn}
=
\mathrm{softmax}\!\left(
g\!\left([\tilde{\mathbf{h}}_m;\tilde{\mathbf{h}}_n]\right)
\right)
\in \mathbb{R}^{K},
\label{eq:fp_prob}
\end{equation}
where $\mathbf{o}_{mn}$ denotes the predicted class distribution over similarity bins and $o_{mn,k}$ is the predicted probability of bin $k$.

The focal loss for a pair $(m,n)$ is defined as:
\begin{equation}
\mathrm{FL}\!\left(\mathbf{o}_{mn}, y_{mn}\right)
=
-\left(1-o_{mn,y_{mn}}\right)^{\gamma}
\log\left(o_{mn,y_{mn}}\right),
\label{eq:fp_focal}
\end{equation}
where $o_{mn,y_{mn}}$ is the predicted probability assigned to the ground-truth bin $y_{mn}$, and $\gamma\ge 0$ is the focusing parameter.

The fingerprint supervision loss is computed over all unordered pairs $(m,n)$ within a mini-batch of size $S$:
\begin{equation}
\mathcal{L}_{\mathrm{fp}}
=
\frac{1}{|\mathcal{P}|}
\sum_{(m,n)\in \mathcal{P}}
\mathrm{FL}\!\left(\mathbf{o}_{mn},\, y_{mn}\right),
\label{eq:fp_loss}
\end{equation}
where $\mathcal{P}=\{(m,n)\mid 1\le m<n\le S\}$ is the set of unique molecule pairs in the batch.

\textbf{Overall first-stage pre-training objective.}
The overall objective of UltraNMR in the first pre-training stage is defined by minimizing a weighted sum of three losses:
\begin{equation}
\mathcal{L}_1
=
\lambda_{\mathrm{mask}} \,
\frac{1}{|\mathcal{S}|}
\sum_{s \in \mathcal{S}}
\mathcal{L}_{\mathrm{mask}}^{(s)}
\;+\;
\lambda_{\mathrm{HC}} \,
\frac{1}{|\mathcal{S}|}
\sum_{s \in \mathcal{S}}
\mathcal{L}_{\mathrm{HC}}^{(s)}
\;+\;
\lambda_{\mathrm{fp}} \,
\mathcal{L}_{\mathrm{fp}},
\label{eq:total_loss}
\end{equation}
where $\mathcal{L}_{\mathrm{mask}}^{(s)}$ and $\mathcal{L}_{\mathrm{HC}}^{(s)}$ are the masked chemical shift prediction loss and proton–carbon correlation loss computed for the $s$-th sample, respectively. The hyperparameters $\lambda_{\mathrm{mask}}$, $\lambda_{\mathrm{HC}}$, and $\lambda_{\mathrm{fp}}$ control the relative contribution of each objective.

\textbf{Isomer contrastive learning.}
In the second stage of UltraNMR pre-training, we aim to enhance the model’s sensitivity to subtle differences in molecular structures~\cite{goldman2023annotating, bushuiev2025self}.
To this end, we construct a specialized isomer dataset by scanning the entire simulated database and grouping NMR spectra according to identical molecular formulas.
For a given input NMR spectrum, its corresponding molecular fingerprint is treated as the positive sample, while the fingerprints of other isomeric molecules sharing the same molecular formula are regarded as negative samples.
Using this isomer dataset, UltraNMR is trained to distinguish the ground-truth molecular fingerprint from those of other isomers via contrastive learning.

Specifically, for a spectrum $s$, we first project its global spectral representation $\tilde{\mathbf{h}}_s$ and molecular fingerprint $\mathbf{p}_s$ into a shared embedding space using two projection networks:
\begin{equation}
\mathbf{v}_s = g_1(\tilde{\mathbf{h}}_s), 
\qquad
\mathbf{u}_s = g_2(\mathbf{p}_s),
\end{equation}
where $g_1(\cdot)$ and $g_2(\cdot)$ are two-layer MLPs.
Both $\mathbf{v}_s$ and $\mathbf{u}_s$ are $\ell_2$-normalized.

The isomer contrastive loss for a spectrum $s$ is defined as:
\begin{equation}
\mathcal{L}_{\mathrm{iso}}^{(s)}
=
- \log
\frac{
\exp\!\left( \mathbf{v}_s^{\top} \mathbf{u}_s / \tau \right)
}{
\exp\!\left( \mathbf{v}_s^{\top} \mathbf{u}_s / \tau \right)
+
\sum_{k=1}^{K}
\exp\!\left( \mathbf{v}_s^{\top} \mathbf{u}_{s,k} / \tau \right)
},
\label{eq:isomer_nce}
\end{equation}
where $\mathbf{u}_{s,k}$ denotes the projected fingerprint embedding of the $k$-th isomeric negative for the spectrum $s$, $K$ is the number of sampled isomeric negatives, and $\tau$ is a temperature hyperparameter set to 0.07.

\textbf{Overall second-stage pre-training objective.}
In the second pre-training stage, UltraNMR is optimized by jointly minimizing the first-stage pre-training objective and the isomer contrastive learning loss over the isomer dataset:
\begin{equation}
\mathcal{L}_{2}
=
(1-\lambda_{\mathrm{iso}}) \mathcal{L}_1
\;+\;
\lambda_{\mathrm{iso}} \,
\frac{1}{|\mathcal{S}|}
\sum_{s \in \mathcal{S}}
\mathcal{L}_{\mathrm{iso}}^{(s)},
\label{eq:stage2_loss}
\end{equation}
where $\lambda_{\mathrm{iso}}$ balances the contribution of isomer contrastive learning.

\subsection{Fine-tuning on Downstream Tasks}
\label{sec:finetuning}

For the superclass classification task, we simply adopt a linear head followed by a softmax function to obtain the predicted probability distribution over superclasses $\hat{\mathbf{y}}_{sc}$:
\begin{equation}
    \hat{\mathbf{y}}_{sc} = \mathrm{softmax}(W_{sc}\tilde{\mathbf{h}} + b_{sc}),
\end{equation}
where $W_{sc} \in \mathbb{R}^{S \times d_{model}}$ and $b_{sc} \in \mathbb{R}^{S}$ are learnable parameters, and $S$ denotes the number of superclasses. 

We employ the cross-entropy loss to fine-tune UltraNMR:
\begin{equation}
\mathcal{L}_{sc} = - \sum_{i=1}^{S} y_{sc,i} \log \hat{y}_{sc,i},
\end{equation}
where $y_{sc}$ is the ground-truth superclass label in one-hot encoding.

For the functional group identification task, we formulate it as a multi-label classification problem. Similar to the superclass classification task, we add a linear head to the UltraNMR framework. 

To alleviate the label imbalance among functional groups, we adopt a Binary Cross-Entropy (BCE) loss combined with focal loss:
\begin{equation}
\mathcal{L}_{\mathrm{fg}}
=
-\frac{1}{M}
\sum_{j=1}^{M}
\left[
\alpha_j y_j
(1-p_j)^{\gamma_j}
\log(p_j)
+
(1-\alpha_j)(1-y_j)
p_j^{\gamma_j}
\log(1-p_j)
\right].
\label{eq:functional_group_loss}
\end{equation}
where $M$ denotes the number of functional groups, 
$y_j \in \{0,1\}$ is the ground-truth indicator for the $j$-th functional group, and 
$\hat{y}_{j}$ is the predicted probability for the $j$-th functional group. 
The hyperparameters $\alpha_j$ and $\gamma_j$ are category-specific weighting and focusing factors determined according to the positive-to-negative sample ratio of each functional group in the training set.

For the \textit{de novo} molecular structure elucidation task, we formulate it as a sequence-to-sequence translation problem based on the learned chemical shift representation $\mathbf{h}^{(L)}$. A Transformer decoder is integrated into the UltraNMR framework to autoregressively generate the molecular representation (e.g., a SMILES string). The model is optimized by minimizing the cross-entropy loss between predicted token distributions and the ground-truth sequence:
\begin{equation}
\mathcal{L}_{dn} = - \sum_{t=1}^{T} \log p(x_t \mid \hat{x}_{<t}, \mathbf{h}^{(L)}),
\end{equation}
where $T$ is the length of the ground-truth sequence and $x_t$ denotes the ground-truth token at position $t$. 

\section{Data and Code availability}
The code for UltraNMR is available at \url{https://github.com/wuycM/UltraNMR}. The checkpoints of UltraNMR are available at \url{https://huggingface.co/milesyc/ultranmr}. The source data of the SimNMR-PubChem Database~\cite{jin2025nmr} can be accessed through \url{https://huggingface.co/datasets/yqj01/SimNMR-PubChem}. The NMRGym~\cite{fang2026nmrgym} dataset can be accessed through \url{https://huggingface.co/datasets/meaw0415/NMRGym}. The processed NMRShiftDB2-2024 data can be obtained via NMRNet~\cite{xu2025toward} at \url{https://zenodo.org/records/19142375}.

%% file: sec/6_supp.tex
\renewcommand\thefigure{\Alph{figure}}
\renewcommand\thetable{\Alph{table}}
\renewcommand\thesection{\Alph{section}}
\setcounter{figure}{0}
\setcounter{table}{0}
\setcounter{section}{0}

\section{Implementation Details}
\label{supp:details}
For simulated NMR data preprocessing, we align the representation of simulated $^{13}$C NMR shifts with that of the experimental data. In experimental spectra, chemically equivalent carbon atoms are typically reported as a single resonance, whereas simulations provide one predicted shift for each individual carbon atom. Therefore, for the simulated $^{13}$C NMR data, shifts corresponding to chemically equivalent carbon atoms are merged into a single value.

For both pre-training and fine-tuning, we implement a stochastic chemical shift augmentation to enhance the model's robustness against experimental variations. Specifically, we randomly apply a global shift bias to both proton ($^{1}$H) and carbon ($^{13}$C) chemical shifts within predefined ranges:
\begin{equation}
\tilde{\delta}_\text{H} = \delta_\text{H} + \epsilon_\text{H}, \quad
\tilde{\delta}_\text{C} = \delta_\text{C} + \epsilon_\text{C},
\end{equation}
where $\epsilon_\text{H} \sim \mathcal{U}(a_\text{H}, b_\text{H})$ and 
$\epsilon_\text{C} \sim \mathcal{U}(a_\text{C}, b_\text{C})$ are uniformly sampled shift biases.

All experiments were conducted on NVIDIA H100 GPUs. Pre-training required 8 hours per epoch using four H100 GPUs, while fine-tuning was completed within a few hours on a single H100 GPU. All neural network architectures were implemented using the PyTorch framework~\cite{paszke2019pytorch}.

\vspace{-1em}
\subsection{Pre-training UltraNMR}
\begin{table}[htbp]
\centering
\caption{Pre-training hyperparameters} 
\label{tab:pt-hyperparameters}
\begin{tabular}{l|c}
\toprule
\textbf{Hyperparameter} & \textbf{Values} \\ \midrule
Learning rate & $1 \cdot 10^{-4}$ \\
Batch size & $256$ \\
Epochs & $6$ \\
Number of warmup steps & $10000$ \\
Dropout & 0.1 \\
Augmentation probability & 0.2 \\
Masked chemical shift ratio & 0.15 \\
Weight decay & $0$ \\
shift bias for ($^{1}$H $a_\text{H}, b_\text{H}$) &(-0.01, 0.01) \\
shift bias for $^{13}$C ($a_\text{C}, b_\text{C}$) &(-0.1, 0.1) \\
($\delta_{min}, \delta_{max})$ for $^{1}$H & (0.01, 16.00) \\
($\delta_{min}, \delta_{max})$ for $^{13}$C & (0.01, 230.00) \\
Bin size for $^{1}$H $\Delta_{\text{H}}$ & 0.01 \\
Bin size for $^{13}$C $\Delta_{\text{C}}$ & 0.1 \\
Standard deviation of the Gaussian kernel for $^{1}$H $\sigma_\text{H}$ & 0.01 \\Standard deviation of the Gaussian kernel for $^{13}$C $\sigma_\text{C}$ & 0.1 \\
Bin size for fingerprint $\Delta_{\text{fp}}$ & 0.005 \\
Focal loss $\gamma$ & 5.0 \\
Masked chemical shift prediction loss weight $\mathcal{L}_{\mathrm{mask}}$ &1.0\\
Proton–carbon correlation loss weight $\mathcal{L}_{\mathrm{HC}}$ & 0.1\\
Fingerprint supervision loss weight $\mathcal{L}_{\mathrm{fp}}$ & 0.1\\
Isomer contrastive loss weight $\mathcal{L}_{\mathrm{iso}}$ & 0.1\\
Number of transformer encoder layers $L$ & $16$ \\
Number of attention heads & $12$ \\
Transformer hidden dimensionality $d_{\text{model}}$ & $768$ \\
Hidden dimensionality of FFN in transformer& $3072$ \\
\bottomrule
\end{tabular}
\end{table}

\newpage
\subsection{Fine-tuning}
During fine-tuning, $^1$H NMR integral information is incorporated by replicating each chemical shift $n$ times, where $n$ denotes the corresponding integral value ($n_H$).
For functional group identification and superclass classification, we define 20 and 74 target labels following the protocols of NMRGym \cite{fang2026nmrgym} and NPClassifier \cite{kim2021npclassifier}, respectively. For the \textit{de novo} molecular structure elucidation task, we employ the ChemBERTa vocabulary \cite{chithrananda2020chemberta} to tokenize SMILES strings. To incorporate molecular formula information, each formula is encoded into a 50-dimensional atom count vector and projected into the model's hidden space using a two-layer MLP. This formula embedding serves as a prefix token, which is stacked at the first row of the input sequence embeddings to guide the structure elucidation process.  During inference, we use beam search to generate top-k candidate SMILES sequences.
\begin{table}[htbp]
\centering
\caption{Fine-tuning hyperparameters} 
\label{tab:ft-hyperparameters}
\begin{tabular}{l|c}
\toprule
\textbf{Hyperparameter} & \textbf{Values} \\ \midrule
Learning rate for functional group identification & $1 \cdot 10^{-4}$ \\
Learning rate for superclass classification & $1 \cdot 10^{-4}$ \\
Learning rate for molecular structure elucidation & $1 \cdot 10^{-5}$ \\
Batch size for functional group identification & 128 \\
Batch size for superclass classification & 128 \\
Batch size for molecular structure elucidation & 64 \\
Epochs & $50$ \\
Dropout & 0.1 \\
Weight decay & $0.01$ \\
Number of transformer decoder layers & 6\\
Length penalty & 1.0 \\
\bottomrule
\end{tabular}
\end{table}

\section{Downstream Dataset Details}
\label{supp:downstream}

We evaluate downstream performance on two experimental NMR datasets: NMRGym and NMRShiftDB. NMRGym~\cite{fang2026nmrgym} is a large-scale benchmark for diverse NMR spectral analysis tasks, built by aggregating experimental NMR spectra from a wide array of public chemical databases. It contains approximately 270K paired $^1$H and $^{13}$C NMR spectra, together with canonicalized molecular structures and peak-level annotations. NMRGym is designed for multiple spectral tasks such as structure elucidation, functional group identification, molecular property prediction, and spectral simulation.

For NMRShiftDB, we follow the processed version used in NMRNet~\cite{xu2025toward}. The data are collected from the nmrshiftdb2 database~\cite{kuhn2015facilitating} in SD-file format, where one-dimensional NMR records are extracted and filtered to remove invalid, empty, unassigned, or out-of-range spectra. Molecules that fail RDKit-based 3D conformer generation are also excluded, and repeated measurements for the same molecule are aggregated by taking the median chemical shift values. The final dataset contains experimentally assigned NMR spectra for multiple nuclei. In our experiments, we retain only the $^1$H and $^{13}$C NMR spectra.
Unlike NMRGym, not all molecules in NMRShiftDB contain paired multimodal spectra; some samples only include a single NMR modality.

\newpage
\section{Quality Evaluation of Simulated NMR Spectra}

To evaluate the quality of simulated NMR spectra, we compare the spectra generated by NMRNet~\cite{xu2025toward} and Gaussian software~\cite{g16} with the experimental spectra in the NMRGym test set. Following previous work~\cite{jin2025nmr}, we adopt two complementary metrics, Vector Similarity and Set Similarity, to measure the alignment between simulated and experimental spectra.

For Vector Similarity, a spectrum is first converted into a continuous signal by Gaussian convolution:
\begin{equation}
g_X(t)=\sum_{i=1}^{n}\exp\!\left(-\frac{(t-x_i)^2}{2\sigma^2}\right),
\end{equation}
where $X=\{x_1,\dots,x_n\}$ denotes the set of chemical shifts and $\sigma$ controls the peak width. The continuous signal is then discretized at 128 uniformly sampled points to form a fixed-dimensional vector:
\begin{equation}
\mathbf{v}_X = [g_X(t_1), g_X(t_2), \dots, g_X(t_{128})].
\end{equation}
The similarity between two spectra $X$ and $Y$ is computed as
\begin{equation}
\mathrm{VecSim}(X,Y)=\frac{\mathbf{v}_X^\top \mathbf{v}_Y}{\|\mathbf{v}_X\|_2 \, \|\mathbf{v}_Y\|_2}.
\end{equation}

For \textit{Set Similarity}, the spectrum is treated as a set of peaks, and similarity is defined by the optimal matching between two peak sets:
\begin{equation}
\mathrm{SetSim}(X,Y)=\frac{1}{\sqrt{mn}} \max_{P}\sum_{(i,j)\in P} f(x_i,y_j),
\end{equation}
where $X=\{x_1,\dots,x_m\}$, $Y=\{y_1,\dots,y_n\}$, $P$ denotes a one-to-one matching between peaks, and the pairwise matching score is defined as
\begin{equation}
f(x,y)=\exp\!\left(-\frac{(x-y)^2}{2\sigma^2}\right).
\end{equation}

Table~\ref{tab:sim_quality} summarizes the results of Gaussian software and NMRNet on the NMRGym test set. Overall, both tools demonstrate strong alignment with experimental spectra, indicating that they can generate high-quality simulated NMR spectra. Gaussian software shows better alignment with experimental spectra for $^1$H simulation, whereas NMRNet achieves slightly stronger performance on $^{13}$C spectra.

\begin{table}[H]
\centering
\caption{Similarity between simulated spectra and experimental spectra on the NMRGym test set.}
\label{tab:sim_quality}
\begin{tabular}{lcccc}
\hline
Method & $^1$H Vec. Sim. & $^1$H Set Sim. & $^{13}$C Vec. Sim. & $^{13}$C Set Sim. \\
\hline
Gaussian software & 0.8987 & 0.8493 & 0.8268 & 0.8997 \\
NMRNet   & 0.8785 & 0.7046 & 0.8492 & 0.9080 \\
\hline
\end{tabular}
\end{table}

\newpage
\section{NMR Spectra of Two Novel Natural Products} \subsection*{Compound 1} 

\textbf{SMILES:} {\texttt{\seqsplit{CC12C(CCC(O3)(CC(C(C)C)=O)C2=CC3=O)C(C)(C)CCC1}}} 

\textbf{Molecular formula:} $\mathrm{C}_{20}\mathrm{H}_{30}\mathrm{O}_{3}$. 

\textbf{$^{1}$H NMR} (400 MHz, CDCl$_3$) $\delta$ 5.68 (s, 1H), 3.19--2.98 (m, 2H), 2.68 (p, $J = 6.9$ Hz, 1H), 2.59 (dt, $J = 9.1, 2.4$ Hz, 1H), 1.86 (dt, $J = 13.7, 4.1$ Hz, 2H), 1.77 (tt, $J = 14.5, 3.9$ Hz, 1H), 1.69--1.57 (m, 2H), 1.56--1.44 (m, 3H), 1.32--1.23 (m, 1H), 1.20 (s, 3H), 1.11 (dd, $J = 6.9, 2.9$ Hz, 6H), 1.02 (td, $J = 8.1, 3.8$ Hz, 1H), 0.94 (d, $J = 3.7$ Hz, 6H). 

\textbf{$^{13}$C NMR} (101 MHz, CDCl$_3$) $\delta$ 210.38, 182.46, 172.23, 111.19, 86.84, 56.39, 46.03, 42.31, 41.72, 40.55, 40.20, 37.25, 34.24, 33.51, 21.71, 19.45, 18.34, 18.07, 17.90, 17.51.

\subsection*{Compound 2} 
\textbf{SMILES:} {\texttt{\seqsplit{C=CCC1=CC=C(C=C1)OC2=C(C3=C(C(CC=C)=C2)C[C@H]([C@@H](O3)C4=CC=C(C(C5=C(C=CC(CC=C)=C5)O)=C4)O)O)O}}}

\textbf{Molecular formula:} $\mathrm{C}_{36}\mathrm{H}_{34}\mathrm{O}_{6}$. 

\textbf{$^{1}$H NMR} (400 MHz, DMSO-$d_6$) $\delta$ 8.48 (s, 1H), 7.15 (d, $J = 1.8$ Hz, 1H), 7.12 (dd, $J = 8.4, 1.8$ Hz, 1H), 7.06 (s, 2H), 6.94 (d, $J = 2.4$ Hz, 1H), 6.90 (dd, $J = 8.4, 1.8$ Hz, 1H), 6.74 (dt, $J = 13.7, 7.9$ Hz, 4H), 6.35 (s, 1H), 5.90 (dddd, $J = 25.4, 18.8, 10.2, 6.5$ Hz, 3H), 5.12--4.91 (m, 7H), 4.76 (d, $J = 7.0$ Hz, 1H), 4.06 (t, $J = 7.4$ Hz, 1H), 3.29 (d, $J = 6.6$ Hz, 2H), 3.25 (d, $J = 6.8$ Hz, 2H), 3.21--3.14 (m, 2H), 2.84 (dd, $J = 16.3, 5.2$ Hz, 1H), 2.61 (dd, $J = 16.3, 7.5$ Hz, 1H). 

\textbf{$^{13}$C NMR} (150 MHz, DMSO-$d_6$) $\delta$ 156.59, 155.30, 154.70, 144.09, 140.73, 138.53, 137.99, 136.82, 136.05, 132.82, 130.89, 129.99, 129.38, 128.60, 128.40, 128.07, 127.78, 126.78, 126.40, 116.59, 116.30, 115.88, 115.66, 115.56, 115.10, 113.77, 81.03, 66.05, 38.94, 38.69, 35.87, 30.28.
\vspace{3em}
\begin{figure}[H]
    \centering
    \includegraphics[width=0.7\textwidth]{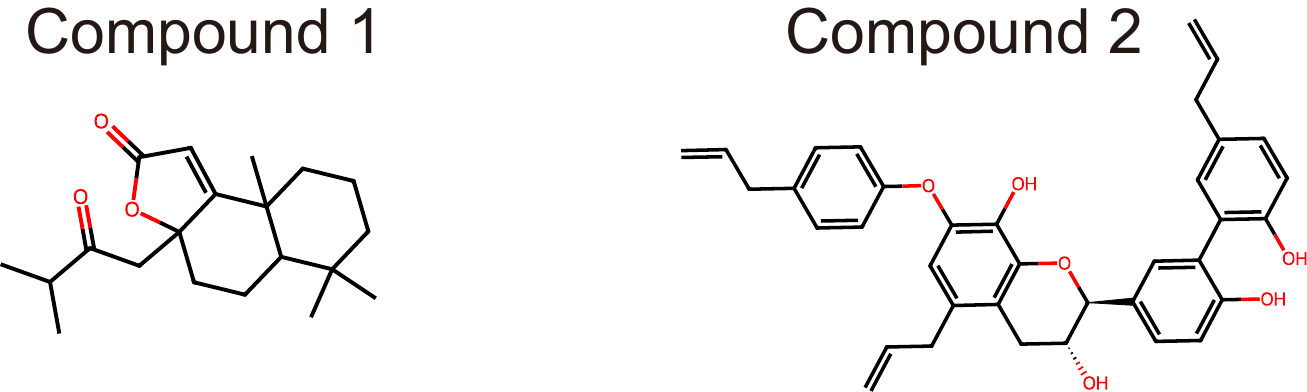}
    \caption{\textbf{Structures of the two novel natural products.} The first compound, named Hypoglycin N, was isolated from \textit{Vitex trifolia} L. var. \textit{simplicifolia} Cham. The second compound, named Magdiligol G, was isolated from \textit{Magnolia officinalis} Rehd. et Wils. var. \textit{biloba} Rehd. et Wils.}
    \label{fig:np}
\end{figure}
\newpage
\section{More Results}

\begin{table}[H]
\centering
\caption{\textit{De novo} molecular structure elucidation performance without molecular formula information.}
\label{tab:nmr_complete_results}
\renewcommand{\arraystretch}{1.2}
\begin{tabular}{llccccccccc}
\toprule
\multirow{2}{*}{\textbf{Dataset}} & \multirow{2}{*}{\textbf{Model}} & \multicolumn{3}{c}{\textbf{Acc (w/o Stereo) \%}} & \multicolumn{3}{c}{\textbf{Acc (w/ Stereo) \%}} & \multicolumn{3}{c}{\textbf{Tanimoto Score}} \\
\cmidrule(lr){3-5} \cmidrule(lr){6-8} \cmidrule(lr){9-11}
& & Top-1 & Top-5 & Top-10 & Top-1 & Top-5 & Top-10 & Top-1 & Top-5 & Top-10 \\
\midrule
\multirow{2}{*}{NMRGym} & UltraNMR & \textbf{7.77} & \textbf{11.45} & \textbf{13.62} & \textbf{3.82} & \textbf{7.63} & \textbf{9.43} & \textbf{0.47} & \textbf{0.52} & \textbf{0.55} \\
                        & NMRMind  & 6.20 & 10.10 & 11.90 & 2.78 & 6.35 & 7.87 & 0.43 & 0.50 & 0.52 \\
\midrule
\multirow{2}{*}{NMRShiftDB} & UltraNMR & \textbf{21.05} & \textbf{29.29} & \textbf{34.29} & \textbf{18.36} & \textbf{26.18} & \textbf{30.74} & \textbf{0.44} & \textbf{0.53} & \textbf{0.57} \\
                            & NMRMind  & 1.90 & 4.10 & 5.40 & 1.59 & 3.55 & 4.63 & 0.16 & 0.23 & 0.26 \\
\bottomrule
\end{tabular}
\end{table}

\begin{figure}[H]
    \centering
    \includegraphics[width=1\textwidth]{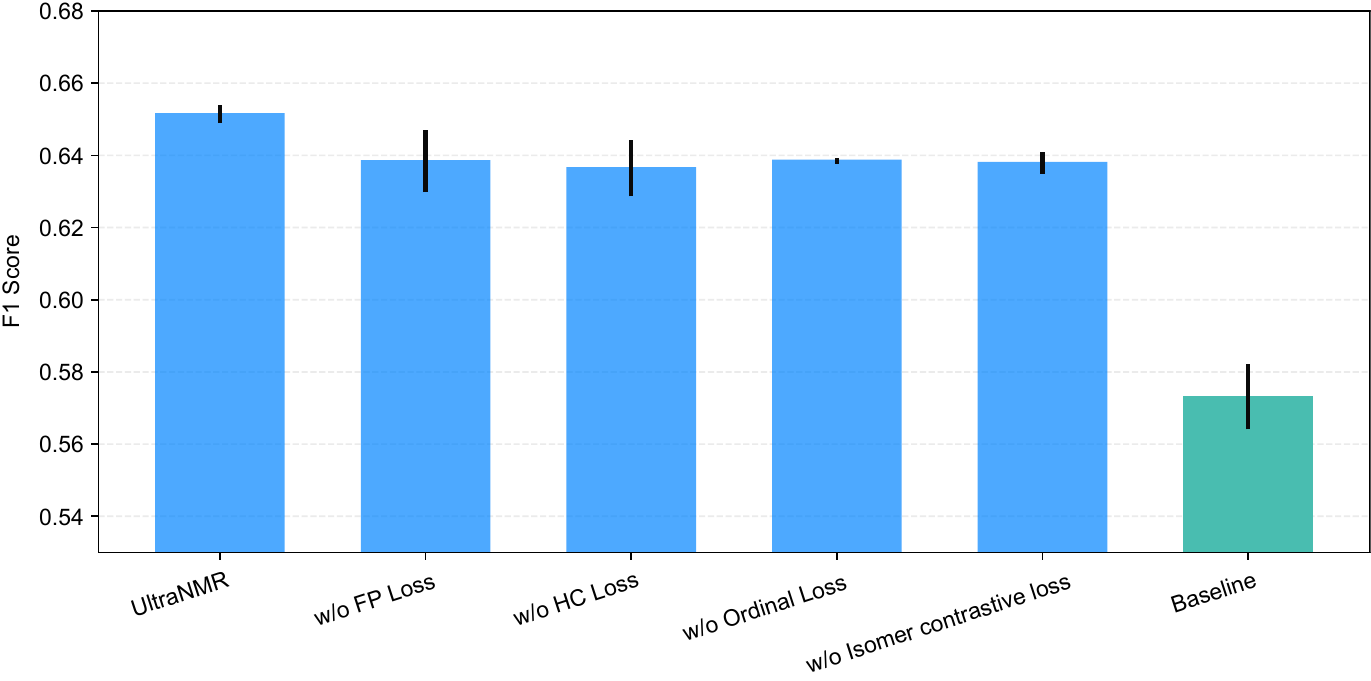}
    \caption{\textbf{Ablation study.} Comparison of the full UltraNMR model against variants missing key pre-training objectives. The evaluated metric is Macro-F1 score for functional group identification on the NMRGym benchmark. w/o FP Loss removes molecular fingerprint supervision that aligns embeddings with structural similarity; w/o HC Loss removes proton–carbon correlation prediction that captures atom-level connectivity; w/o Isomer contrastive loss removes isomer contrastive learning that enhances the model’s sensitivity to subtle differences in molecular structures; and w/o Ordinal Loss replaces the distance-aware ordinal regression task with a standard classification task for masked chemical shift prediction.}
    \label{fig:ablation}
\end{figure}

\newpage

\begin{figure}[H]
    \centering
    \includegraphics[scale=0.75]{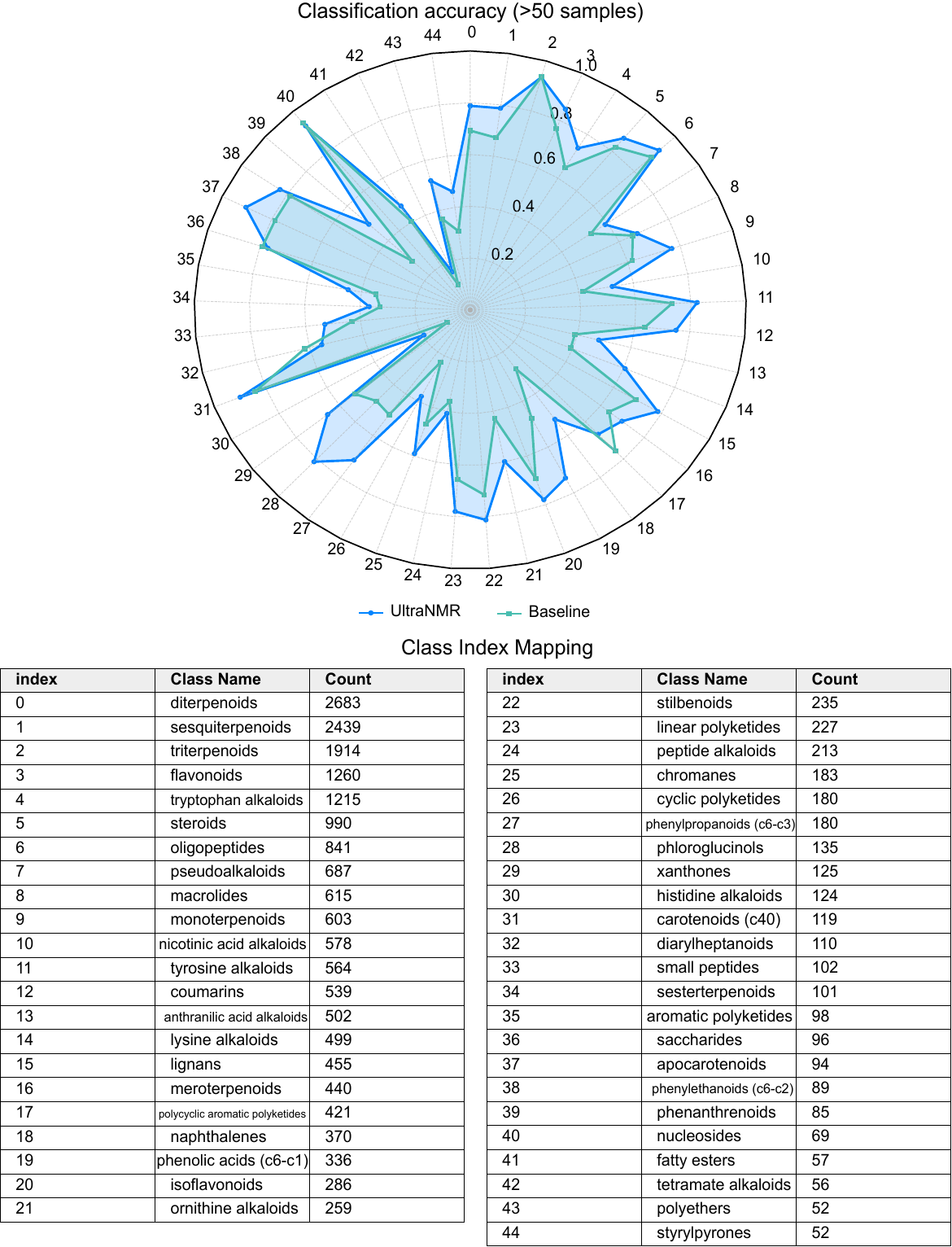}
    \caption{\textbf{Superclass classification performance on NMRGym.} Comparison of classification accuracy between UltraNMR and the baseline, evaluated on all classes with more than 50 samples.}
    \label{fig:superclass}
\end{figure}

\begin{figure}[H]
    \centering
    \includegraphics[width=1\textwidth]{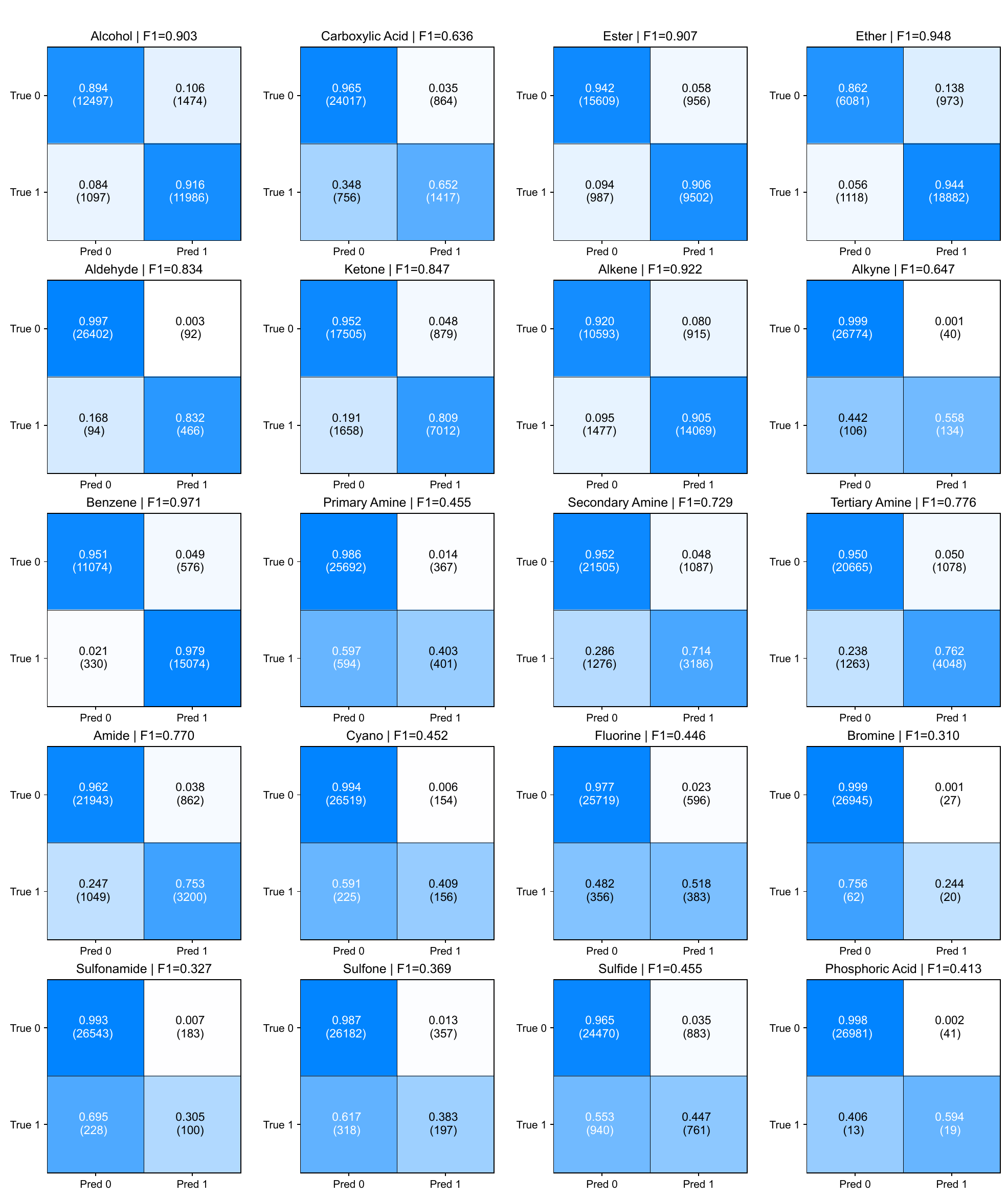}
    \caption{\textbf{Confusion matrices for each functional group on NMRGym.} Cell values represent normalized ratios and absolute sample counts (in parentheses).}
    \label{fig:fg}
\end{figure}

\newpage
\section{Full Numerical Results}

\begin{table}[htbp]
\centering
\caption{Accuracy and Recall metrics of library search on NMRGym.}
\label{tab:acc_recall_corrected}
\small
\setlength{\tabcolsep}{3.5pt}
\begin{tabular}{l ccc ccc ccc ccc}
\toprule
\multirow{2.5}{*}{\textbf{Method}} & \multicolumn{3}{c}{\textbf{Acc. w Stereo (\%) $\uparrow$}} & \multicolumn{3}{c}{\textbf{Rec. w Stereo (\%) $\uparrow$}} & \multicolumn{3}{c}{\textbf{Acc. w/o Stereo (\%) $\uparrow$}} & \multicolumn{3}{c}{\textbf{Rec. w/o Stereo (\%) $\uparrow$}}\\
\cmidrule(lr){2-4} \cmidrule(lr){5-7} \cmidrule(lr){8-10} \cmidrule(lr){11-13}
& T-1 & T-5 & T-10 & T-1 & T-5 & T-10 & T-1 & T-5 & T-10 & T-1 & T-5 & T-10 \\
\midrule
Gaussian & 14.20 & 44.54 & 56.97 & 24.00 & 66.94 & 82.85 & 41.54 & 56.47 & 61.83 & 52.75 & 71.70 & 78.51 \\
UltraNMR & \textbf{15.51} & \textbf{48.78} & \textbf{61.83} & \textbf{25.31} & \textbf{70.45} & \textbf{86.17} & \textbf{43.17} & \textbf{58.45} & \textbf{63.88} & \textbf{54.81} & \textbf{74.22} & \textbf{81.11} \\
\bottomrule
\end{tabular}
\end{table}

\begin{table}[htbp]
\centering
\caption{Tanimoto similarity scores of library search on NMRGym.}
\label{tab:tanimoto_results}
\small
\begin{tabular}{l ccc}
\toprule
\textbf{Method} & \textbf{Top-1 Mean $\uparrow$} & \textbf{Top-5 Mean $\uparrow$} & \textbf{Top-10 Mean $\uparrow$} \\
\midrule
Gaussian & 0.61 & 0.72 & 0.76 \\
UltraNMR & \textbf{0.65} & \textbf{0.76} & \textbf{0.79} \\
\bottomrule
\end{tabular}
\end{table}

\begin{table}[H]
\centering
\caption{Full test metrics of functional group identification on NMRGym.}
\label{tab:model_performance}
\begin{tabular}{lcccc}
\toprule
\textbf{Method} & \textbf{Accuracy (\%) $\uparrow$} & \textbf{Macro-F1 (\%) $\uparrow$} & \textbf{Macro-Recall (\%) $\uparrow$} \\
\midrule
XGBoost      & 45.68 & 55.47 & 45.02 \\
NMR2Struct   & 36.25 & 55.44 & 58.62\\
Baseline     & 34.28 & 57.31 & 59.95  \\
Gaussian     & 39.97 & 58.82 & 59.55  \\
\textbf{UltraNMR} & \textbf{50.20} & \textbf{65.14} & \textbf{64.14}\\
\bottomrule
\end{tabular}
\end{table}

\begin{table}[H]
\centering
\caption{Full test metrics of \textit{de novo} molecular structure elucidation on NMRGym.}
\label{tab:nmrgym_results}
\small
\begin{tabular}{l ccc ccc ccc}
\toprule
\multirow{2.5}{*}{\textbf{Method}} & \multicolumn{3}{c}{\textbf{Acc. w/ Stereo (\%) $\uparrow$}} & \multicolumn{3}{c}{\textbf{Acc. w/o Stereo (\%) $\uparrow$}} & \multicolumn{3}{c}{\textbf{Tanimoto Score $\uparrow$}} \\
\cmidrule(lr){2-4} \cmidrule(lr){5-7} \cmidrule(lr){8-10}
& Top-1 & Top-5 & Top-10 & Top-1 & Top-5 & Top-10 & Top-1 & Top-5 & Top-10 \\
\midrule
Baseline & 0.34 & 0.81 & 1.13 & 0.92 & 1.66 & 2.10 & 0.36 & 0.39 & 0.40 \\
NMRMind  & 4.67 & 10.17 & 12.42 & 10.10 & 15.70 & 18.30 & 0.46 & 0.54 & 0.56 \\
UltraNMR & \textbf{6.39} & \textbf{12.39} & \textbf{14.68} & \textbf{12.83} & \textbf{18.45} & \textbf{21.03} & \textbf{0.51} & \textbf{0.57} & \textbf{0.60} \\
\bottomrule
\end{tabular}
\end{table}

\begin{table}[H]
\centering
\caption{Full test metrics of \textit{de novo} molecular structure elucidation on NMRShiftDB.}
\label{tab:nmrshiftdb_results}
\small
\begin{tabular}{l ccc ccc ccc}
\toprule
\multirow{2.5}{*}{\textbf{Method}} & \multicolumn{3}{c}{\textbf{Acc. w/ Stereo (\%) $\uparrow$}} & \multicolumn{3}{c}{\textbf{Acc. w/o Stereo (\%) $\uparrow$}} & \multicolumn{3}{c}{\textbf{Tanimoto Score $\uparrow$}} \\
\cmidrule(lr){2-4} \cmidrule(lr){5-7} \cmidrule(lr){8-10}
& Top-1 & Top-5 & Top-10 & Top-1 & Top-5 & Top-10 & Top-1 & Top-5 & Top-10 \\
\midrule
Baseline & 12.34 & 25.09 & 29.35 & 15.61 & 29.01 & 33.37 & 0.43 & 0.53 & 0.57 \\
NMRMind  & 9.18 & 19.85 & 24.21 & 11.90 & 23.75 & 28.55 & 0.33 & 0.47 & 0.52 \\
UltraNMR & \textbf{33.15} & \textbf{45.97} & \textbf{49.79} & \textbf{38.37} & \textbf{51.48} & \textbf{55.37} & \textbf{0.60} & \textbf{0.69} & \textbf{0.73} \\
\bottomrule
\end{tabular}
\end{table}

%% file: ref.bib
@article{southey2023introduction,
  title={Introduction to small molecule drug discovery and preclinical development},
  author={Southey, Michelle WY and Brunavs, Michael},
  journal={Frontiers in Drug Discovery},
  volume={3},
  pages={1314077},
  year={2023},
  publisher={Frontiers Media SA}
}

@inproceedings{xie2022much,
  title={How much of the chemical space has been explored? selecting the right exploration measure for drug discovery},
  author={Xie, Yutong and Xu, Ziqiao and Ma, Jiaqi and Mei, Qiaozhu},
  booktitle={ICML 2022 2nd AI for Science Workshop},
  year={2022}
}

@article{kirkpatrick2004chemical,
  title={Chemical space.},
  author={Kirkpatrick, Peter and Ellis, Clare},
  journal={Nature},
  volume={432},
  number={7019},
  pages={823--824},
  year={2004},
  publisher={Nature Publishing Group}
}

@article{kim2025pubchem,
  title={PubChem 2025 update},
  author={Kim, Sunghwan and Chen, Jie and Cheng, Tiejun and Gindulyte, Asta and He, Jia and He, Siqian and Li, Qingliang and Shoemaker, Benjamin A and Thiessen, Paul A and Yu, Bo and others},
  journal={Nucleic acids research},
  volume={53},
  number={D1},
  pages={D1516--D1525},
  year={2025},
  publisher={Oxford University Press}
}

@article{bishop2023robust,
  title={Robust automated backbone triple resonance NMR assignments of proteins using Bayesian-based simulated annealing},
  author={Bishop, Anthony C and Torres-Montalvo, Gloris{\'e} and Kotaru, Sravya and Mimun, Kyle and Wand, A Joshua},
  journal={Nature Communications},
  volume={14},
  number={1},
  pages={1556},
  year={2023},
  publisher={Nature Publishing Group UK London}
}

@article{giraudeau2017challenges,
  title={Challenges and perspectives in quantitative NMR},
  author={Giraudeau, Patrick},
  journal={Magnetic Resonance in Chemistry},
  volume={55},
  number={1},
  pages={61--69},
  year={2017},
  publisher={Wiley Online Library}
}

@inproceedings{alberts2023learning,
  title={Learning the language of NMR: structure elucidation from NMR spectra using transformer models},
  author={Alberts, Marvin and Zipoli, Federico and Vaucher, Alain},
  booktitle={AI for Accelerated Materials Design-NeurIPS 2023 Workshop},
  year={2023}
}

@article{tan2025transformer,
  title={A transformer based generative chemical language AI model for structural elucidation of organic compounds},
  author={Tan, Xiaofeng},
  journal={Journal of cheminformatics},
  volume={17},
  number={1},
  pages={103},
  year={2025},
  publisher={Springer}
}

@article{xue2025nmrmind,
  title={NMRMind: A Transformer-Based Model Enabling the Elucidation from Multidimensional NMR to Structures},
  author={Xue, Xi and Sun, Hanyu and Sun, Jingying and Patiny, Luc and Liu, Xiangying and Chen, Kai and Yan, Jingjie and Li, Liangning and Liu, Xue and Xu, Shu and others},
  journal={Analytical Chemistry},
  volume={97},
  number={41},
  pages={22603--22614},
  year={2025},
  publisher={ACS Publications}
}

@article{bushuiev2025self,
  title={Self-supervised learning of molecular representations from millions of tandem mass spectra using DreaMS},
  author={Bushuiev, Roman and Bushuiev, Anton and Samusevich, Raman and Brungs, Corinna and Sivic, Josef and Pluskal, Tom{\'a}{\v{s}}},
  journal={Nature Biotechnology},
  pages={1--11},
  year={2025},
  publisher={Nature Publishing Group US New York}
}

@article{hu2024accurate,
  title={Accurate and efficient structure elucidation from routine one-dimensional nmr spectra using multitask machine learning},
  author={Hu, Frank and Chen, Michael S and Rotskoff, Grant M and Kanan, Matthew W and Markland, Thomas E},
  journal={ACS Central Science},
  volume={10},
  number={11},
  pages={2162--2170},
  year={2024},
  publisher={ACS Publications}
}

@article{xiongatomic,
  title={Atomic diffusion models for small molecule structure elucidation from nmr spectra},
  author={Xiong, Ziyu and Zhang, Yichi and Alauddin, Foyez and Cheng, Chu Xin and An, Joon and Seyedsayamdost, Mohammad and Zhong, Ellen},
  journal={Advances in Neural Information Processing Systems},
  volume={38},
  pages={115995--116031},
  year={2026}
}

@article{yang2025diffnmr,
  title={DiffNMR: Diffusion Models for Nuclear Magnetic Resonance Spectra Elucidation},
  author={Yang, Qingsong and Wu, Binglan and Liu, Xuwei and Chen, Bo and Li, Wei and Long, Gen and Chen, Xin and Xiao, Mingjun},
  journal={Materials Futures},
  year={2025}
}

@article{jin2025nmr,
  title={NMR-Solver: automated structure elucidation via large-scale spectral matching and physics-guided fragment optimization},
  author={Jin, Yongqi and Wang, Jun-Jie and Xu, Fanjie and Ji, Xiaohong and Gao, Zhifeng and Zhang, Linfeng and Ke, Guolin and Zhu, Rong and E, Weinan},
  journal={Nature Communications},
  year={2026},
  publisher={Nature Publishing Group UK London}
}

@article{xu2025toward,
  title={Toward a unified benchmark and framework for deep learning-based prediction of nuclear magnetic resonance chemical shifts},
  author={Xu, Fanjie and Guo, Wentao and Wang, Feng and Yao, Lin and Wang, Hongshuai and Tang, Fujie and Gao, Zhifeng and Zhang, Linfeng and E, Weinan and Tian, Zhong-Qun and others},
  journal={Nature Computational Science},
  pages={1--9},
  year={2025},
  publisher={Nature Publishing Group US New York}
}

@article{simeoni2025dinov3,
  title={Dinov3},
  author={Sim{\'e}oni, Oriane and Vo, Huy V and Seitzer, Maximilian and Baldassarre, Federico and Oquab, Maxime and Jose, Cijo and Khalidov, Vasil and Szafraniec, Marc and Yi, Seungeun and Ramamonjisoa, Micha{\"e}l and others},
  journal={arXiv preprint arXiv:2508.10104},
  year={2025}
}

@article{zhou2023foundation,
  title={A foundation model for generalizable disease detection from retinal images},
  author={Zhou, Yukun and Chia, Mark A and Wagner, Siegfried K and Ayhan, Murat S and Williamson, Dominic J and Struyven, Robbert R and Liu, Timing and Xu, Moucheng and Lozano, Mateo G and Woodward-Court, Peter and others},
  journal={Nature},
  volume={622},
  number={7981},
  pages={156--163},
  year={2023},
  publisher={Nature Publishing Group UK London}
}

@article{vaswani2017attention,
  title={Attention is all you need},
  author={Vaswani, Ashish and Shazeer, Noam and Parmar, Niki and Uszkoreit, Jakob and Jones, Llion and Gomez, Aidan N and Kaiser, {\L}ukasz and Polosukhin, Illia},
  journal={Advances in neural information processing systems},
  volume={30},
  year={2017}
}

@inproceedings{devlin2019bert,
  title={Bert: Pre-training of deep bidirectional transformers for language understanding},
  author={Devlin, Jacob and Chang, Ming-Wei and Lee, Kenton and Toutanova, Kristina},
  booktitle={Proceedings of the 2019 conference of the North American chapter of the association for computational linguistics: human language technologies, volume 1 (long and short papers)},
  pages={4171--4186},
  year={2019}
}

@inproceedings{radford2021learning,
  title={Learning transferable visual models from natural language supervision},
  author={Radford, Alec and Kim, Jong Wook and Hallacy, Chris and Ramesh, Aditya and Goh, Gabriel and Agarwal, Sandhini and Sastry, Girish and Askell, Amanda and Mishkin, Pamela and Clark, Jack and others},
  booktitle={International conference on machine learning},
  pages={8748--8763},
  year={2021},
  organization={PmLR}
}

@article{raffel2020exploring,
  title={Exploring the limits of transfer learning with a unified text-to-text transformer},
  author={Raffel, Colin and Shazeer, Noam and Roberts, Adam and Lee, Katherine and Narang, Sharan and Matena, Michael and Zhou, Yanqi and Li, Wei and Liu, Peter J},
  journal={Journal of machine learning research},
  volume={21},
  number={140},
  pages={1--67},
  year={2020}
}

@article{newman2020natural,
  title={Natural products as sources of new drugs over the nearly four decades from 01/1981 to 09/2019},
  author={Newman, David J and Cragg, Gordon M},
  journal={Journal of natural products},
  volume={83},
  number={3},
  pages={770--803},
  year={2020},
  publisher={ACS Publications}
}

@article{fukuto2012small,
  title={Small molecule signaling agents: the integrated chemistry and biochemistry of nitrogen oxides, oxides of carbon, dioxygen, hydrogen sulfide, and their derived species},
  author={Fukuto, Jon M and Carrington, Samantha J and Tantillo, Dean J and Harrison, Jason G and Ignarro, Louis J and Freeman, Bruce A and Chen, Andrew and Wink, David A},
  journal={Chemical research in toxicology},
  volume={25},
  number={4},
  pages={769--793},
  year={2012},
  publisher={ACS Publications}
}

@article{qiu2023small,
  title={Small molecule metabolites: discovery of biomarkers and therapeutic targets},
  author={Qiu, Shi and Cai, Ying and Yao, Hong and Lin, Chunsheng and Xie, Yiqiang and Tang, Songqi and Zhang, Aihua},
  journal={Signal Transduction and Targeted Therapy},
  volume={8},
  number={1},
  pages={132},
  year={2023},
  publisher={Nature Publishing Group UK London}
}

@article{duhrkop2019sirius,
  title={SIRIUS 4: a rapid tool for turning tandem mass spectra into metabolite structure information},
  author={D{\"u}hrkop, Kai and Fleischauer, Markus and Ludwig, Marcus and Aksenov, Alexander A and Melnik, Alexey V and Meusel, Marvin and Dorrestein, Pieter C and Rousu, Juho and B{\"o}cker, Sebastian},
  journal={Nature methods},
  volume={16},
  number={4},
  pages={299--302},
  year={2019},
  publisher={Nature Publishing Group US New York}
}

@article{bruguiere202113c,
  title={13C NMR dereplication using MixONat software: a practical guide to decipher natural products mixtures},
  author={Brugui{\`e}re, Antoine and Derbr{\'e}, S{\'e}verine and Br{\'e}ard, Dimitri and Tomi, F{\'e}lix and Nuzillard, Jean-Marc and Richomme, Pascal},
  journal={Planta Medica},
  volume={87},
  number={12/13},
  pages={1061--1068},
  year={2021},
  publisher={Georg Thieme Verlag KG}
}

@article{yang2021cross,
  title={Cross-modal retrieval between 13C NMR spectra and structures for compound identification using deep contrastive learning},
  author={Yang, Zhuo and Song, Jianfei and Yang, Minjian and Yao, Lin and Zhang, Jiahua and Shi, Hui and Ji, Xiangyang and Deng, Yafeng and Wang, Xiaojian},
  journal={Analytical Chemistry},
  volume={93},
  number={50},
  pages={16947--16955},
  year={2021},
  publisher={ACS Publications}
}

@article{wang2025nmrexp,
  title={NMRexp: A database of 3.3 million experimental NMR spectra},
  author={Wang, Jun-Jie and Jin, Yongqi and Zhi, Chen-Yu and Liu, Yu-Jie and Huang, Xu-Hao and Xu, Fanjie and Ji, Xiaohong and Fang, Xi and Tao, Haoyi and E, Weinan and others},
  journal={Scientific Data},
  volume={12},
  number={1},
  pages={1954},
  year={2025},
  publisher={Nature Publishing Group UK London}
}

@inproceedings{chen2020simple,
  title={A simple framework for contrastive learning of visual representations},
  author={Chen, Ting and Kornblith, Simon and Norouzi, Mohammad and Hinton, Geoffrey},
  booktitle={International conference on machine learning},
  pages={1597--1607},
  year={2020},
  organization={PmLR}
}

@inproceedings{assran2023self,
  title={Self-supervised learning from images with a joint-embedding predictive architecture},
  author={Assran, Mahmoud and Duval, Quentin and Misra, Ishan and Bojanowski, Piotr and Vincent, Pascal and Rabbat, Michael and LeCun, Yann and Ballas, Nicolas},
  booktitle={Proceedings of the IEEE/CVF Conference on Computer Vision and Pattern Recognition},
  pages={15619--15629},
  year={2023}
}

@inproceedings{he2020momentum,
  title={Momentum contrast for unsupervised visual representation learning},
  author={He, Kaiming and Fan, Haoqi and Wu, Yuxin and Xie, Saining and Girshick, Ross},
  booktitle={Proceedings of the IEEE/CVF conference on computer vision and pattern recognition},
  pages={9729--9738},
  year={2020}
}

@article{tancik2020fourier,
  title={Fourier features let networks learn high frequency functions in low dimensional domains},
  author={Tancik, Matthew and Srinivasan, Pratul and Mildenhall, Ben and Fridovich-Keil, Sara and Raghavan, Nithin and Singhal, Utkarsh and Ramamoorthi, Ravi and Barron, Jonathan and Ng, Ren},
  journal={Advances in neural information processing systems},
  volume={33},
  pages={7537--7547},
  year={2020}
}

@inproceedings{rahaman2019spectral,
  title={On the spectral bias of neural networks},
  author={Rahaman, Nasim and Baratin, Aristide and Arpit, Devansh and Draxler, Felix and Lin, Min and Hamprecht, Fred and Bengio, Yoshua and Courville, Aaron},
  booktitle={International conference on machine learning},
  pages={5301--5310},
  year={2019},
  organization={PMLR}
}

@article{ba2016layer,
  title={Layer normalization},
  author={Ba, Jimmy Lei and Kiros, Jamie Ryan and Hinton, Geoffrey E},
  journal={arXiv preprint arXiv:1607.06450},
  year={2016}
}

@inproceedings{he2016deep,
  title={Deep residual learning for image recognition},
  author={He, Kaiming and Zhang, Xiangyu and Ren, Shaoqing and Sun, Jian},
  booktitle={Proceedings of the IEEE conference on computer vision and pattern recognition},
  pages={770--778},
  year={2016}
}

@article{hendrycks2016gaussian,
  title={Gaussian Error Linear Units (Gelus)},
  author={Hendrycks, D},
  journal={arXiv preprint arXiv:1606.08415},
  year={2016}
}

@inproceedings{tan2016age,
  title={Age estimation based on a single network with soft softmax of aging modeling},
  author={Tan, Zichang and Zhou, Shuai and Wan, Jun and Lei, Zhen and Li, Stan Z},
  booktitle={Asian Conference on Computer Vision},
  pages={203--216},
  year={2016},
  organization={Springer}
}

@article{wang2025survey,
  title={A Survey on Ordinal Regression: Applications, Advances and Prospects},
  author={Wang, Jinhong and Chen, Jintai and Liu, Jian and Tang, Dongqi and Chen, Danny Z and Wu, Jian},
  journal={arXiv preprint arXiv:2503.00952},
  year={2025}
}

@article{xiong2025supervised,
  title={Supervised Contrastive Learning Leads to More Reasonable Spectral Embeddings},
  author={Xiong, Peng and Xu, Hongtao and Zheng, Haoran},
  journal={Analytical Chemistry},
  volume={97},
  number={37},
  pages={20137--20146},
  year={2025},
  publisher={ACS Publications}
}

@article{goldman2023annotating,
  title={Annotating metabolite mass spectra with domain-inspired chemical formula transformers},
  author={Goldman, Samuel and Wohlwend, Jeremy and Stra{\v{z}}ar, Martin and Haroush, Guy and Xavier, Ramnik J and Coley, Connor W},
  journal={Nature Machine Intelligence},
  volume={5},
  number={9},
  pages={965--979},
  year={2023},
  publisher={Nature Publishing Group UK London}
}

@inproceedings{lin2017focal,
  title={Focal loss for dense object detection},
  author={Lin, Tsung-Yi and Goyal, Priya and Girshick, Ross and He, Kaiming and Doll{\'a}r, Piotr},
  booktitle={Proceedings of the IEEE international conference on computer vision},
  pages={2980--2988},
  year={2017}
}

@article{kim2021npclassifier,
  title={NPClassifier: a deep neural network-based structural classification tool for natural products},
  author={Kim, Hyun Woo and Wang, Mingxun and Leber, Christopher A and Nothias, Louis-F{\'e}lix and Reher, Raphael and Kang, Kyo Bin and Van Der Hooft, Justin JJ and Dorrestein, Pieter C and Gerwick, William H and Cottrell, Garrison W},
  journal={Journal of natural products},
  volume={84},
  number={11},
  pages={2795--2807},
  year={2021},
  publisher={ACS Publications}
}

@article{fang2026nmrgym,
  title={NMRGym: A Comprehensive Benchmark for Nuclear Magnetic Resonance Based Molecular Structure Elucidation},
  author={Fang, Zheng and Yang, Chen and Yu, Hai-tao and Luo, Haoming and He, Haitao and Xie, Jiaqing and Yang, Zhuo and Xia, Jun},
  journal={arXiv preprint arXiv:2601.15763},
  year={2026}
}

@article{mcinnes2018umap,
  title={Umap: Uniform manifold approximation and projection for dimension reduction},
  author={McInnes, Leland and Healy, John and Melville, James},
  journal={arXiv preprint arXiv:1802.03426},
  year={2018}
}

@article{kuhn2015facilitating,
  title={Facilitating quality control for spectra assignments of small organic molecules: nmrshiftdb2--a free in-house NMR database with integrated LIMS for academic service laboratories},
  author={Kuhn, Stefan and Schl{\"o}rer, Nils E},
  journal={Magnetic Resonance in Chemistry},
  volume={53},
  number={8},
  pages={582--589},
  year={2015},
  publisher={Wiley Online Library}
}

@article{paszke2019pytorch,
  title={Pytorch: An imperative style, high-performance deep learning library},
  author={Paszke, Adam and Gross, Sam and Massa, Francisco and Lerer, Adam and Bradbury, James and Chanan, Gregory and Killeen, Trevor and Lin, Zeming and Gimelshein, Natalia and Antiga, Luca and others},
  journal={Advances in neural information processing systems},
  volume={32},
  year={2019}
}

@article{chithrananda2020chemberta,
  title={ChemBERTa: large-scale self-supervised pretraining for molecular property prediction},
  author={Chithrananda, Seyone and Grand, Gabriel and Ramsundar, Bharath},
  journal={arXiv preprint arXiv:2010.09885},
  year={2020}
}

@article{yang2021predicting,
  title={Predicting chemical shifts with graph neural networks},
  author={Yang, Ziyue and Chakraborty, Maghesree and White, Andrew D},
  journal={Chemical science},
  volume={12},
  number={32},
  pages={10802--10809},
  year={2021},
  publisher={Royal Society of Chemistry}
}

@article{gerrard2020impression,
  title={IMPRESSION--prediction of NMR parameters for 3-dimensional chemical structures using machine learning with near quantum chemical accuracy},
  author={Gerrard, Will and Bratholm, Lars A and Packer, Martin J and Mulholland, Adrian J and Glowacki, David R and Butts, Craig P},
  journal={Chemical science},
  volume={11},
  number={2},
  pages={508--515},
  year={2020},
  publisher={Royal Society of Chemistry}
}

@misc{g16,
author={M. J. Frisch and G. W. Trucks and H. B. Schlegel and G. E. Scuseria and M. A. Robb and J. R. Cheeseman and G. Scalmani and V. Barone and G. A. Petersson and H. Nakatsuji and X. Li and M. Caricato and A. V. Marenich and J. Bloino and B. G. Janesko and R. Gomperts and B. Mennucci and H. P. Hratchian and J. V. Ortiz and A. F. Izmaylov and J. L. Sonnenberg and D. Williams-Young and F. Ding and F. Lipparini and F. Egidi and J. Goings and B. Peng and A. Petrone and T. Henderson and D. Ranasinghe and V. G. Zakrzewski and J. Gao and N. Rega and G. Zheng and W. Liang and M. Hada and M. Ehara and K. Toyota and R. Fukuda and J. Hasegawa and M. Ishida and T. Nakajima and Y. Honda and O. Kitao and H. Nakai and T. Vreven and K. Throssell and Montgomery, {Jr.}, J. A. and J. E. Peralta and F. Ogliaro and M. J. Bearpark and J. J. Heyd and E. N. Brothers and K. N. Kudin and V. N. Staroverov and T. A. Keith and R. Kobayashi and J. Normand and K. Raghavachari and A. P. Rendell and J. C. Burant and S. S. Iyengar and J. Tomasi and M. Cossi and J. M. Millam and M. Klene and C. Adamo and R. Cammi and J. W. Ochterski and R. L. Martin and K. Morokuma and O. Farkas and J. B. Foresman and D. J. Fox},
title={Gaussian˜16 {R}evision {C}.01},
year={2016},
note={Gaussian Inc. Wallingford CT}
}

@article{kim2023deepsat,
  title={DeepSAT: learning molecular structures from nuclear magnetic resonance data},
  author={Kim, Hyun Woo and Zhang, Chen and Reher, Raphael and Wang, Mingxun and Alexander, Kelsey L and Nothias, Louis-F{\'e}lix and Han, Yoo Kyong and Shin, Hyeji and Lee, Ki Yong and Lee, Kyu Hyeong and others},
  journal={Journal of Cheminformatics},
  volume={15},
  number={1},
  pages={71},
  year={2023},
  publisher={Springer}
}

@inproceedings{10.24963/ijcai.2025/1160,
author = {Guo, Kehan and Shen, Yili and Gonzalez-Montiel, Gisela Abigail and Huang, Yue and Zhou, Yujun and Surve, Mihir and Guo, Zhichun and Das, Payel and Chawla, Nitesh V. and Wiest, Olaf and Zhang, Xiangliang},
title = {Artificial intelligence in spectroscopy: advancing chemistry from prediction to generation and beyond},
year = {2025},
isbn = {978-1-956792-06-5},
url = {https://doi.org/10.24963/ijcai.2025/1160},
doi = {10.24963/ijcai.2025/1160},
articleno = {1160},
numpages = {10},
location = {Montreal, Canada},
series = {IJCAI '25}
}

@article{luo2025deep,
  title={Deep learning and its applications in nuclear magnetic resonance spectroscopy},
  author={Luo, Yao and Zheng, Xiaoxu and Qiu, Mengjie and Gou, Yaoping and Yang, Zhengxian and Qu, Xiaobo and Chen, Zhong and Lin, Yanqin},
  journal={Progress in Nuclear Magnetic Resonance Spectroscopy},
  volume={146},
  pages={101556},
  year={2025},
  publisher={Elsevier}
}

@article{li2022identifying,
  title={Identifying molecular functional groups of organic compounds by deep learning of NMR data},
  author={Li, Chongcan and Cong, Yong and Deng, Weihua},
  journal={Magnetic Resonance in Chemistry},
  volume={60},
  number={11},
  pages={1061--1069},
  year={2022},
  publisher={Wiley Online Library}
}

@article{lee2025machine,
  title={Machine-learning approach to identify organic functional groups from ft-ir and nmr spectral data},
  author={Lee, Gwanho and Shim, Hyekyoung and Cho, Juhyun and Choi, Sang-Il},
  journal={ACS omega},
  volume={10},
  number={12},
  pages={12717--12723},
  year={2025},
  publisher={ACS Publications}
}

@article{kuhn2022pilot,
  title={A pilot study for fragment identification using 2D NMR and deep learning},
  author={Kuhn, Stefan and Tumer, Eda and Colreavy-Donnelly, Simon and Moreira Borges, Ricardo},
  journal={Magnetic Resonance in Chemistry},
  volume={60},
  number={11},
  pages={1052--1060},
  year={2022},
  publisher={Wiley Online Library}
}

@article{martinez2020prediction,
  title={Prediction of natural product classes using machine learning and 13C NMR spectroscopic data},
  author={Martinez-Trevino, Saul H and Uc-Cetina, Victor and Fern{\'a}ndez-Herrera, Mar{\'\i}a A and Merino, Gabriel},
  journal={Journal of Chemical Information and Modeling},
  volume={60},
  number={7},
  pages={3376--3386},
  year={2020},
  publisher={ACS Publications}
}

@article{cui2024scgpt,
  title={scGPT: toward building a foundation model for single-cell multi-omics using generative AI},
  author={Cui, Haotian and Wang, Chloe and Maan, Hassaan and Pang, Kuan and Luo, Fengning and Duan, Nan and Wang, Bo},
  journal={Nature methods},
  volume={21},
  number={8},
  pages={1470--1480},
  year={2024},
  publisher={Nature Publishing Group US New York}
}
